\documentclass[12pt,a4paper]{article}
\pdfoutput=1
\usepackage{jheppub}

\usepackage{amsmath,amssymb,epsfig,bm,amsfonts,yfonts,mathrsfs}

\relax
\title{Isotropization from Color Field Condensate in heavy ion collisions}

\author[a]{Stefan Floerchinger} 
\author[b]{and Christof Wetterich}

\affiliation[a]{Physics Department, Theory Unit, CERN, CH-1211 Gen\`eve 23, Switzerland}

\affiliation[b]{Institut f\"ur Theoretische Physik, Philosophenweg 16, 69120 Heidelberg, Germany}

\emailAdd{Stefan.Floerchinger@cern.ch}
\emailAdd{C.Wetterich@thphys.uni-heidelberg.de}

\abstract{
The expanding fireball shortly after a heavy ion collision may be qualitatively described by a condensate of color fields or gluons which is analogous to Bose-Einstein-condensation for massive bosonic particles. This condensate is a transient non-equilibrium phenomenon and breaks Lorentz-boost symmetry. The dynamics of color field condensates involves collective excitations and is rather different from the perturbative scattering of gluons. In particular, it provides for an efficient mechanism to render the local pressure approximately isotropic after a short time of 0.2 fm/c. We suggest that an isotropic color field condensate may play a central role for a simple description of prethermalization and isotropization in the early stages of the collision.}

\begin{document}
\maketitle
\section{Introduction}
Hydrodynamics can give a reasonable description of the dynamics of heavy ion collisions if it applies already in rather early stages of the fireball. Fast interactions have to create an approximately isotropic local pressure within a rather short time of about 0.5 fm/c. If the strong gauge coupling $\alpha_S(p)=g^2(p)/4\pi$ determines the strength of the interactions of gluons and quarks with momenta of the order of the effective temperature, the time scales estimated from perturbation theory \cite{Baier:2000sb} are much too long in order to account for the early validity of hydrodynamics. This discrepancy has been a puzzle for many years, see \cite{Berges:2012ks} for a survey of recent works on this topic.

The early validity of hydrodynamics does not require that a thermodynamic equilibrium state is approximately reached. An effective temperature $T$, related to the kinetic energy of the particles, may be established by a very rapid prethermalization \cite{Berges:2004ce}. Furthermore, isotropisation of the pressure can occur on time scales much shorter than the ones needed for thermal equilibration \cite{Berges:2005ai}. In a particle picture an approximate isotropisation of the pressure requires scattering processes that can change the directions of momenta at a sufficient rate. Many mechanisms have been proposed to achieve a fast isotropisation, for example plasma instabilities
\cite{Romatschke:2003ms,Romatschke:2005pm,Romatschke:2006nk,Arnold:2003rq,Arnold:2004ti,Arnold:2005vb,Arnold:2005ef,Randrup:2003cw,Mrowczynski:2005ki,Dumitru:2006pz,Berges:2008zt,Bodeker:2005nv,Rebhan:2004ur,Rebhan:2005re,Romatschke:2006wg,Rebhan:2008uj,Ipp:2010uy,Gelis:2013rba,Epelbaum:2013waa,Kurkela:2011ub,Kurkela:2011ti,Berges:2011sb,Attems:2012js,Berges:2012iw,Berges:2012cj,Berges:2013eia,Berges:2013fga}, 
chaotic behavior 
\cite{Muller:1992iw,Biro:1994sh,Kunihiro:2010tg,Nishiyama:2010mn} 
or strong coupling behavior as simulated by AdS/CFT models \cite{Chesler:2008hg,Chesler:2009cy,Kovchegov:2009du,Balasubramanian:2010ce,Beuf:2009cx,Heller:2011ju}
, but none seems completely satisfactory so far. In this note we propose that the dynamics of a gluon field condensate could provide a simple mechanism for rapid isotropization.

More generally, we propose that a large number density of gluons may be the essential ingredient to induce effective interactions that are substantially stronger than the ones expected in perturbative QCD and could help to achieve early isotropisation and approximate validity of hydrodynamics. In other words, collective effects of many gluons are much more efficient for isotropization than the individual scattering of single gluons.

\begin{figure}
\begin{center}
\includegraphics[width=0.5\textwidth]{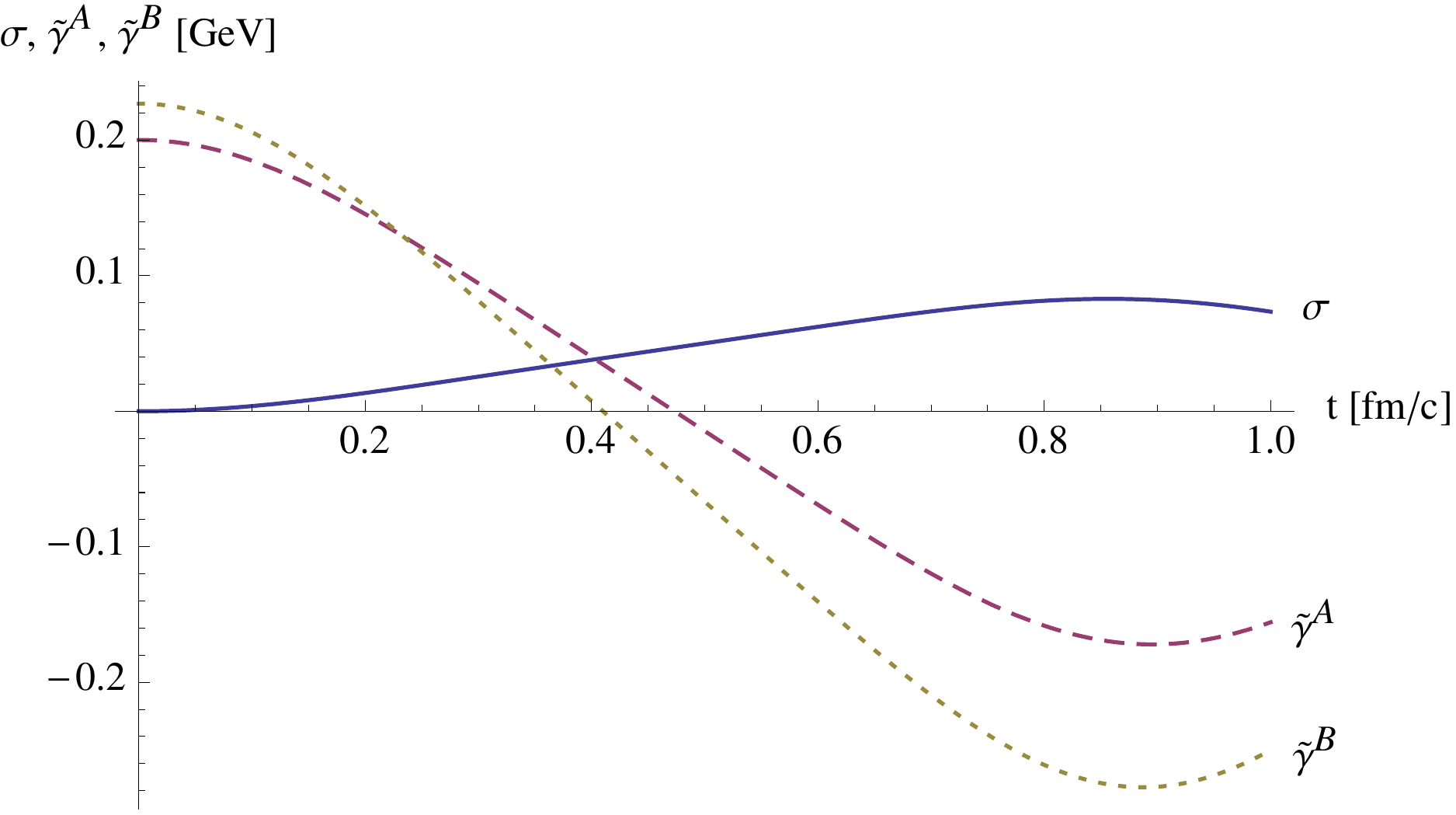}
\caption{Time evolution of the isotropic condensate $\sigma(t)$ (solid line) and cylindrical condensates $\tilde \gamma^A(t)$ (dashed line) and $\tilde \gamma^B(t)$ (dotted line) in a homogeneous situation with initial condition $\dot\sigma(0) = \dot{\tilde \gamma}^{A}(0)= \dot{\tilde \gamma}^{B}(0)=0$. The initial values $\tilde \gamma^A(0)$ and $\tilde \gamma^B(0)$ have been chosen such that the longitudinal pressure $p_l$ vanishes initially. The isotropic condensate vanishes initially, $\sigma(0)=0$, and is generated by the time evolution.}
\label{fig7}
\end{center}
\end{figure}

\begin{figure}
\begin{center}
\includegraphics[width=0.5\textwidth]{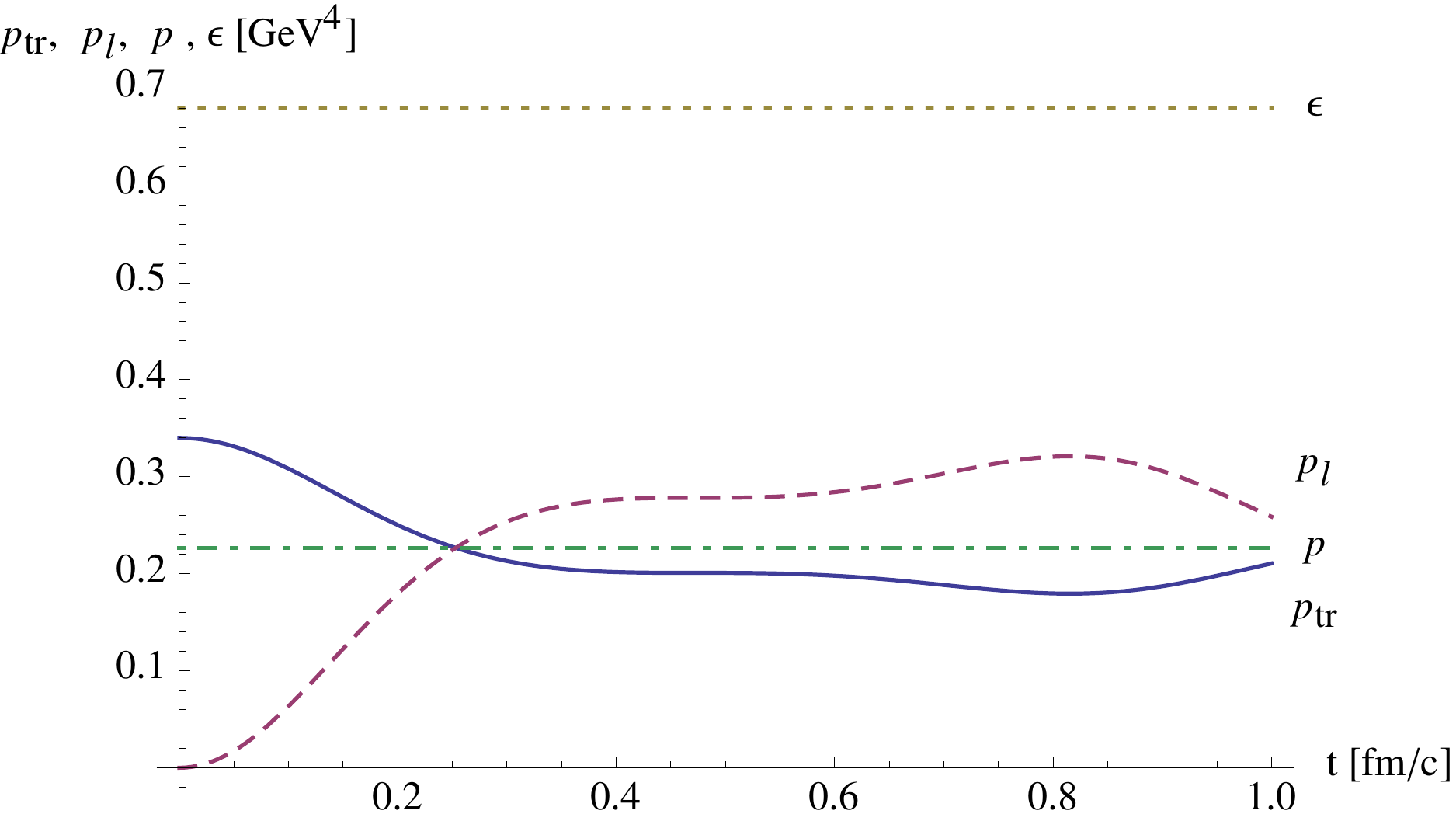}
\caption{Time evolution of the transverse pressure $p_{tr}(t)$ (solid line),  longitudinal pressure $p_l(t)$ (dashed line) corresponding to the time dependent condensates as shown in Fig.\ \ref{fig7}. We also indicate the energy density $\epsilon(t)$ (dotted line) and the averaged pressure $p(t) = (2 p_{tr}(t)+p_l(t))/3 = \epsilon(t)/3$ (dashed-dotted line).}
\label{fig8}
\end{center}
\end{figure}

The easiest description of a large gluon number may be a condensate of the gluon field, corresponding to a ``classical gluon field''. An isotropic energy momentum tensor can be achieved by a field \cite{Reuter:1994yq} of the form $\bold{A}_0=0$ (in Weyl or temporal gauge) and
\begin{equation}
\bold{A}_1=  \sigma
\begin{pmatrix} 0, & 0, & 0 \\
0, & 0, & i  \\
0, & -i , & 0 
\end{pmatrix}, \quad
\bold{A}_2= \sigma 
\begin{pmatrix} 0, & 0, & -i \\
0, & 0, & 0 \\
i, & 0, & 0 
\end{pmatrix},\quad 
\bold{A}_3= \sigma 
\begin{pmatrix} 
0, & i, & 0 \\
-i, & 0, & 0 \\
0, & 0, & 0 
\end{pmatrix}.
\label{eq:singletconfigurationexplicit}
\end{equation}
We develop a simple picture where an initial field with cylindrical symmetry and vanishing longitudinal pressure develops rapidly into the isotropic field condensate \eqref{eq:singletconfigurationexplicit}, thereby realizing an isotropic pressure. This is demonstrated in Fig. \ref{fig7} which shows how an initially vanishing isotropic field $\sigma$ increases rapidly, generated by substantial fields with cylindrical symmetry. 
In Fig.\ \ref{fig8} we plot the corresponding values of longitudinal and transverse pressure. Already at a very short time of about $0.2 \, \text{fm}$ an almost isotropic pressure is realized. At this time the isotropic condensate $\sigma$ is still rather small, demonstrating that an approximate isotropization of pressure can also be due to the time evolution of fields that are not rotation invariant. At $t$ somewhat larger than 0.4 fm/c the isotropic condensate has begun to dominate. 
For later times the evolution of the condensates and pressure components may deviate substantially from Figs.\ \ref{fig7} and \ref{fig8} as we discuss below. 
It seems unlikely, however, that this destroys the approximate isotropy of the pressure that has been achieved at this stage. In any case, we expect that the isotropic condensate will start to decrease at some moment. This may be due to the oscillatory tendency of classical solutions, due to the production of incoherent particles, or simply due to the decrease of the condensate energy density as a result of the expansion of the fireball. An isotropic condensate may finally decay either directly into hadrons or into the quasiparticles of the quark  gluon plasma. In both cases the pressure remains isotropic. It is even possible than an isotropic color field condensate survives in an approximate thermal equilibrium state of a quark gluon plasma \cite{Reuter:1994yq}. For the equilibrium state it will not play a leading role, however.

In the presence of a color field condensate also the properties of excitations or quasiparticles are strongly modified. For example the effective cubic coupling between three gluons becomes $g^2 \sigma$ instead of $g\, p_\text{Temp}$ for perturbative QCD, with $p_\text{Temp}$ a characteristic momentum associated to the particle distribution function realized for a corresponding effective temperature. This will lead to an enhanced interaction strength for $\sigma > p_\text{Temp}/g$. The scattering amplitude for $2\to 2$ scattering with exchanged momentum $\sim p_\text{Temp}$ is boosted to $\sim g^4 \sigma^2/p_\text{Temp}^2$ instead of $g^2$. 

For an order of magnitude estimate of a possible condensate one may consider a time of $0.3 \; \text{fm/c}$ and a temperature scale $T=0.61 \text{GeV}$, characteristic for heavy ion collisions at the LHC. We take $g=\sqrt{4\pi \alpha_S(p_\text{Temp})} \approx 2.5 $ for a momentum scale $p_\text{Temp}\approx 1.6 \, T \approx 1\, \text{GeV} $, as given by the maximum of the black body spectrum $p^3 n_B(p,T)$ for massless particles. Equating the pressure for the condensate $p=2 g^2 \sigma^4$ (see below) with the pressure of a realistic QCD equation of state $p(T)\approx 0.68 \, \text{GeV}^4$ one obtains $\sigma\approx 0.5 \, \text{GeV}$, with lower values if only a fraction of the pressure is attributed to the condensate. For these values $g^2 \sigma \approx 3.1\, \text{GeV}$ is of the order of $g p_\text{Temp} \approx 2.5 \, \text{GeV}$ and even somewhat larger. As the temperature scale decreases in later stages of the collision $\sigma$ and $p_\text{Temp}$ both decrease.

The ``color field condensate'' or ``gluon field condensate'' considered in this paper corresponds to a non-zero value of the vector potential $A_\mu$, not to be confounded with the ``gluon condensate'' or ``glueball condensate'' corresponding to non-zero value of the invariant $\langle F^{\mu\nu} F_{\mu\nu}\rangle$. It is considered here as a transient phenomenon of the non-equilibrium situation in the early stages of the fireball and it vanishes in the vacuum. 

In the thermodynamic limit a spatially constant condensate corresponds to macroscopic occupation numbers of gluons with vanishing momentum. For a finite particle number, finite volume and a non-equilibrium situation the distinction between a condensate and large occupation numbers for gluons with small momenta is not so clear. Still, a description in terms of condensates or equivalent classical gluon fields may be a good guide for phenomena where a large number density of low-momentum gluons plays a crucial role.
The possibility of condensates forming in heavy ion collisions has been investigated recently in approaches different from ours \cite{Blaizot:2011xf,Epelbaum:2011pc,Berges:2012us,Berges:2012ev,Kurkela:2012hp,Blaizot:2012qd,Blaizot:2013lga}. We consider our picture of a color field condensate as a simple way of catching some relevant features of non-equilibrium physics with high gluon occupation numbers or distributions of gluon fields. 

In this paper we concentrate on rotation symmetric gluon field condensates as discussed in ref. \ \cite{Reuter:1994yq}. For colored fields as the vector potential $\bold{A}_\mu^a$ the rotations are realized as a combination of usual space-rotations with suitable gauge transformations. All gauge singlets transform under rotations in the standard way. This holds, in particular, for the energy momentum tensor. We emphasize that a locally isotropic condensate is not in contradiction to the reduced symmetry of the fireball in a real scattering event - very similar to the notion of a locally isotropic pressure. 

Before starting, it may be useful to place the proposed color field condensate in the more general context of Bose-Einstein condensation. Bose-Einstein condensation governs the (many body) ground state at zero temperature and non-zero density for non-relativistic massive bosons. Two aspects distinguish this phenomenon from other condensation phenomena in relativistic quantum field theory such as the chiral or Higgs field condensate. The first is that particle number conservation plays an important role for Bose-Einstein condensation while this is not the case for the relativistic condensates. The second is that the latter have a higher degree of symmetry, being invariant under Lorentz transformations. We discuss first the aspect of symmetries and comment on the role of conservation laws later on.

The chiral and Higgs condensates are vacuum phenomena and therefore symmetric with respect to the full Poincar\'e group of rotations, Lorentz boosts and space-time translations. For the traditional Bose-Einstein condensation this group is broken down to a smaller subgroup. The highest degree of symmetry is maintained for the (somewhat hypothetical) case of a BEC that is homogeneous, isotropic and stationary in time. In that case the symmetry under rotations as well as space- and time translations is preserved. Lorentz boost invariance is broken, however. This can directly be seen from the state itself or by taking into consideration that there is a singled out reference frame where the fluid formed by the BEC is at rest. 
Another way to formulate this is in terms of the fluid velocity. To a Bose-Einstein Condensate in equilibrium one can associate a constant fluid (four-) velocity $u^\mu$. In the fluid rest frame it is of the form $u^\mu=(1,0,0,0)$. (More generally, one may define the fluid velocity in the Landau frame such that $T^{0\mu} = T^{\mu 0} \sim u^\mu$.) In the following we will work in the fluid rest frame unless mentioned otherwise.

While the situation is rather simple and well understood for scalar fields it becomes somewhat more complicated for other bosonic fields. For example, the electromagnetic field transforms as a vector under Lorentz transformations so that a Bose-Einstein condensation that conserves rotations and translations is only conceivable for its temporal component, which, however, is a gauge degree of freedom. Indeed, a non-zero value $A_0$ that is constant in space and time does not have a physical significance and can be gauged away. For massive vector fields the situation is different. As an example, normal nuclear matter contains a condensate of (the temporal component of) the isospin singlet vector boson $\omega_\mu$. Similar to a BEC of massive scalars this would not be possible at vanishing density due to Lorentz-boost symmetry. However, at non-zero density it is allowed by the symmetries. Since $\omega_\mu$ is a real singlet field the condensate does not break any continuous internal symmetry and therefore does not lead to the presence of gapless Goldstone bosons or superfluidity.

Conventional non-relativistic Bose-Einstein Condensates correspond to macroscopic fields or expectation values for complex scalar fields. A global U(1) symmetry implies via Noethers theorem the conservation of particle number. A positive value of the associated chemical potential implies at zero temperature and for a repulsive interaction spontaneous symmetry breaking and superfluidity.

For more unconventional types of Bose-Einstein condensation such as a BEC of photons in a cavity \cite{BECinaCavity} particle number may not be exactly conserved. However, excitations in a particular mode may decay only very slowly or an external source may provide them in a steady way so that a Bose-Einstein type condensate can occur as a metastable phenomenon or as a non-equilibrium steady state. The color field condensate discussed in this paper shows some analogies to the photon condensate. The main distinctions are its occurrence only as a transient phenomenon and the possible realization of three-dimensional rotation symmetry while the BEC of photons \cite{BECinaCavity} is two-dimensional.
\section{Isotropic and cylindrical color field condensate}
\label{sec:IsotropicColorFieldCondensate}

A macroscopic value or condensate of the gluon field $(A_\mu)_{mn} = A_\mu^a (t^a)_{mn}$ necessarily breaks both Lorentz and color rotation symmetry. (The generators $t^a$ are given by the Gell-Mann matrices, $t^a=\frac{1}{2}\lambda_a$. We work with signature $(-,+++)$.) Moreover, a macroscopic value of $u^\mu (A_\mu)_{mn} = (A_0)_{mn}$, which would preserve spatial rotation symmetry in the fluid rest frame, does not have physical meaning. It is gauge-equivalent to $(A_\mu)_{mn}=0$. By choosing Weyl gauge one can actually always fix $u^\mu (A_\mu)_{mn} = (A_0)_{mn}=0$. A first guess may therefore suggest that for a non-vanishing gluon field condensate of the space components $(A_j)_{mn}$ also ordinary rotation invariance has to be broken. There is, however, another possibilty: one can define a modified rotation symmetry which combines ordinary spatial rotations with a rotation in color space (a gauge transformation). For any observations or experimental tests these modified rotations do not differ from ordinary rotations since all gauge invariant objects (gauge symmetry singlets) transform in the normal way.

In a group theoretic language a rotation invariant situation can be realized by embedding the rotation group $O(3)$ (or its double coverage $SU(2)$) into the gauge group $SU(3)$. There are two inequivalent possible embeddings of this type that may be used for our purpose \cite{Reuter:1994yq}. In the first case the Lie algebra of $SU(2)$ is spanned by the Gell-Mann matrices $\lambda_2$, $\lambda_5$ and $\lambda_7$. The singlet corresponds to the configuration
\begin{equation}
(A_j)_{mn} = i \sigma \epsilon_{jmn}.
\label{eq:singletconfiguraioncase1}
\end{equation}
With respect to the combined rotations the octet of ``space-like'' gluons $(A_j)_{mn}$ contains a singlet while the eight components of the temporal part $(A_0)_{mn}$ transform as $\bold{5} + \bold{3}$. We will concentrate on this choice. We mention, however, the possibility of an alternative embedding where the Lie algebra of $SU(2)$ is spanned by $\lambda_1$, $\lambda_2$ and $\lambda_3$. In this case one has two possible singlets, a non-trivial one in the spatial part of the gauge field,
\begin{equation}
(A_j)_{mn} = \frac{1}{\sqrt{2}} \sigma (\lambda_j)_{mn}
\label{eq:2.2}
\end{equation}
and a trivial one in the temporal part, $(A_0)_{mn}=\sigma^\prime \lambda_8$. The temporal part transforms as $\bold{3}+\bold{2}+\bold{2}+\bold{1}$. A residual U(1) symmetry is preserved by the configuration \eqref{eq:2.2}.

Concentrating on the choice \eqref{eq:singletconfiguraioncase1} the explicit form of the gluon field condensate is the one of Eq.\ \eqref{eq:singletconfigurationexplicit}. The combined rotations act on the gauge fields $(A_0)_{mn}$ and $(A_j)_{mn}$ infinitesimally as
\begin{equation}
\begin{split}
\delta (A_0)_{mn} = & i \alpha^k (\hat t^k)_{mp} (A_0)_{pn} +  i \alpha^k (\hat t^k)_{np} (A_0)_{mp},\\
\delta (A_j)_{mn} = & i \alpha^k (\hat t^k)_{jp} (A_p)_{mn} +i \alpha^k (\hat t^k)_{mp} (A_j)_{pn} +  i \alpha^k (\hat t^k)_{np} (A_j)_{mp},
\end{split}
\label{eq:2.3}
\end{equation}
where $\vec \alpha$ parametrizes the infinitesimal rotations and $(\hat t^k)_{mn}=-i \epsilon_{kmn}$ are the generators of the SO(3) Lie algebra. For homogeneous fields it is straightforward to verify the invariance of the configuration \eqref{eq:singletconfiguraioncase1}. For inhomogeneous fields one has to accompany the transformation \eqref{eq:2.3} by a suitable infinitesimal transformation of the space coordinates.

The field configurations which involve only the singlet $\sigma$ have isotropic energy momentum tensor, $T^{\mu\nu} = \text{diag}(\epsilon, p, p, p)$, with relativistic conformal equation of state $\epsilon=3 p$. For our normalization of $\sigma$  the pressure obeys
\begin{equation}
p = 2 g^2 \sigma^4.
\end{equation}
In our simple picture the isotropic condensate $\sigma$ becomes the central carrier of isotropic pressure at an early stage of the fireball.

The initial gluon field configurations created directly at the collision (possibly associated to a color glass condensate) have a different symmetry. Originally, the longitudinal component of the pressure $p_l$ may be very small, reflecting approximate free streaming in the direction of the collision axis. Even negative values of $p_l$ could be realized, for example associated to an ``initial color field condensate''. 
For this initial stage the three dimensional rotation symmetry is broken down to cylindrical symmetry, i.\ e.\ rotation symmetry in the two-dimensional plane of $x_1$ and $x_2$, supplemented by rotations of $180$ degree around the $x_1$ axis, $(x_1,x_2,x_3) \to (x_1,-x_2,-x_3)$ (or, equivalently, around the $x_2$ axis, $(x_1,x_2,x_3) \to (-x_1, x_2,-x_3)$). Field configurations that respect these symmetries are of the form
\begin{equation}
\begin{split}
\bold{A}_1 = & 
\begin{pmatrix}
0, & 0, & 0 \\
0, & 0,&  - 3 \tilde \gamma^A + i \tilde \gamma^B + i \sigma \\
0, & -3 \tilde \gamma^A - i \tilde \gamma^B - i \sigma, & 0
\end{pmatrix},\\
\bold{A}_2 = & 
\begin{pmatrix}
0, & 0, & 3 \tilde \gamma^A - i \tilde \gamma^B - i \sigma \\
0, & 0,& 0  \\
3 \tilde \gamma^A + i \tilde \gamma^B + i \sigma, & 0, & 0
\end{pmatrix},\\
\bold{A}_3 = & 
\begin{pmatrix}
0, & - 2 i \tilde \gamma^B + i \sigma, & 0, & 0 \\
2 i \tilde \gamma^B - i \sigma, & 0,& 0 \\
0, & 0, & 0
\end{pmatrix}.
\label{eq:Acylindric}
\end{split}
\end{equation}
We associate the initial stage with the cylindrical condensates $\tilde \gamma^A$ and $\tilde \gamma^B$. For $\tilde \gamma^A=0$ the condensate is symmetric with respect to the discrete charge symmetry $\mathsf{C}$ discussed at the end of section \ref{sec:Decompositionofgaugefield}, while for $\tilde \gamma^B=\sigma=0$ the discrete $\mathsf{CP}$-symmetry is preserved. The initial energy-momentum tensor is, in general, far from isotropic.

For the classical Yang-Mills action the contribution of the gluon field to the energy-momentum tensor reads
\begin{equation}
T^{\mu\nu} = 2 \text{tr} \; \bold{F}^{\rho\mu} \bold{F}_\rho^{\;\;\nu} - \frac{1}{2} g^{\mu\nu} \; \text{tr} \; \bold{F}^{\alpha\beta} \bold{F}_{\alpha\beta}.
\end{equation}
The trace vanishes, $T^\mu_\mu = 0$; we neglect here the trace anomaly associated to running couplings.
In an isotropic situation the energy momentum tensor takes the form
\begin{equation}
T^{\mu\nu} = (\epsilon+p) u^{\mu} u^{\nu} + p g^{\mu\nu},
\end{equation}
with $u^{\mu}=(1,0,0,0)$ in the rest frame. Its expression in terms of ``electric'' and ``magnetic'' gluon fields, $\bold{E}_j=\bold{F}_{0j}$, $\bold{B}_j=-\frac{1}{2}\epsilon_{jkl} \bold{F}_{kl}$, reads
\begin{equation}
\begin{split}
\epsilon = & \text{tr}\;( \bold{E}_j \bold{E}_j + \bold{B}_j \bold{B}_j),\\
p = & \frac{1}{3}  \text{tr}\;( \bold{E}_j \bold{E}_j + \bold{B}_j \bold{B}_j).
\end{split}
\label{eq:epsiolnpisotropic}
\end{equation}
This implies the equation of state $\epsilon = 3 p$. 
For a field configuration with a possibly time dependent but spatially constant isotropic field $\sigma$ one finds
\begin{equation}
\epsilon =  6 (\partial_0 \sigma)^2 + 6 g^2 \sigma^4, \quad \quad p=\epsilon/3.
\label{eq:nergypressuresigma}
\end{equation}

In the more general situation of only cylindrical symmetry one has (in the fluid rest frame)
\begin{equation}
T^{\mu\nu} = \text{diag} (\epsilon, p_{tr}, p_{tr}, p_l)
\end{equation}
where $\epsilon$ is given as in \eqref{eq:epsiolnpisotropic} and
\begin{equation}
\begin{split}
p_{tr} = & \text{tr} \;( 
\bold{E}_3^2 
+ \bold{B}_3^2),\\
p_l = & \text{tr} \;( \bold{E}_1^2 + \bold{E}_2^2 - \bold{E}_3^2 + \bold{B}_1^2 + \bold{B}_2^2 - \bold{B}_3^2).
\end{split}
\end{equation}
The ``equation of state'' becomes now $\epsilon=2 p_{tr} + p_l$, and we may define the average pressure $p=(2p_{tr}+p_l)/3 = \epsilon/3$. For the gauge field \eqref{eq:Acylindric} we find (using the relations \eqref{eq:definitionE}, \eqref{eq:definitionB})
\begin{equation}
\begin{split}
\epsilon = & 36 \, (\partial_0 \tilde \gamma^A)^2 + 12 \, (\partial_0 \tilde \gamma^B)^2 + 6 \, (\partial_0 \sigma)^2 \\
& +  6\,  g^2 \left( 3 (\tilde \gamma^A)^2 + 2 (\tilde \gamma^B)^2 + (\tilde \gamma^B - \sigma)^2 \right) \left( 9 (\tilde \gamma^A)^2 + (\tilde \gamma^B + \sigma)^2 \right), \\
p_{tr} = & 2 \left(2 \partial_0 \tilde \gamma^B - \partial_0 \sigma \right)^2 + 2 g^2 \left( 9 (\tilde \gamma^A)^2 + (\tilde \gamma^B + \sigma)^2 \right)^2,\\
p_l = & \epsilon - 2 p_{tr}.
\end{split} 
\end{equation}
It is interesting that $\epsilon$ and $p_{tr}$ are positive while $p_l$ can also be negative. This allows us to choose initial configurations with vanishing or negative longitudinal pressure.
\section{Time evolution of condensates}
\label{sec:TimeEvolutionOfCondensate}
Even if an isotropic condensate is created in the early stages of an heavy ion collision it may decay too fast to be relevant for the process of isotropization. We therefore want first to get an idea how long a gluon field condensate $\sigma$ could remain an important ingredient. For this purpose we concentrate on the time evolution of $\sigma$ and neglect other possible color field condensates as $\tilde \gamma^{A,B}$.

An estimate of the time evolution of an isotropic condensate contains two aspects. The first concerns a completely rotation symmetric situation. In this case one can restrict the discussion to finding an (approximate) field equation for $\sigma$ and its solution. The second issue concerns the role of the non-singlet fluctuations. For realistic collision geometries gradient terms will excite such non-singlet fluctuations even if they were negligible at a given moment. We expect non-singlet fluctuations to be present at all stages of the collision. We therefore have to investigate the stability of the proposed rotation symmetric condensate with respect to non-symmetric excitations. This will be done in sects.\ \ref{sec:Excitations} and \ref{sec:StabilityAnalysis}.

Finding the relevant field equation which describes the time evolution of $\sigma$ is a rather complicated task. For a non-equilibrium situation with high particle density it is equivalent to finding the non-equilibrium effective action \cite{Wetterich:1996ap} from which it can be derived by variation. The non-equilibrium effective action may differ substantially from the quantum effective action which describes the vacuum and its excitations at equilibrium. We only know that it needs to be symmetric under gauge transformations as well as other symmetries relevant for the non-equilibrium situation. For a first start we will, nevertheless, approximate the relevant non-equilibrium effective action by the standard yang-Mills action of QCD
\begin{equation}
\mathscr{L}= - \frac{1}{2} \text{tr}\; \bold{F}_{\mu\nu} \bold{F}^{\mu\nu},
\label{eq:YMLagrangian}
\end{equation}
with
\begin{equation}
\bold{F}_{\mu\nu} = \partial_\mu \bold{A}_\nu - \partial_\nu \bold{A}_\mu - i g [\bold{A}_\mu\bold{A}_\nu].
\end{equation}
(We neglect quark degrees of freedom in this exploratory study.) We are aware that additional terms omitted in \eqref{eq:YMLagrangian} could well lead to an additional stabilization (or destabilization) of the condensate. For example, it is easy to construct an (euclidean) effective action for which the configuration \eqref{eq:singletconfiguraioncase1} constitutes the minimum and is therefore stable.

With the approximation \eqref{eq:YMLagrangian} the evolution equation for the homogeneous field $\sigma$ can be obtained by inserting the gauge field configuration \eqref{eq:singletconfiguraioncase1} into the Yang-Mills field equation
\begin{equation}
[\bold{D}_\nu, \bold{F}^{\mu\nu}] = 0, \quad \bold{D}_\nu = \partial_\nu - i g \bold{A}_\nu.
\end{equation}
Equivalently, we can insert \eqref{eq:singletconfiguraioncase1} in the action \eqref{eq:YMLagrangian}, 
\begin{equation}
\mathscr{L}_\sigma = 6 (\partial_0 \sigma)^2 - 4 \partial_j \sigma \partial_j \sigma - 6 g^2 \sigma^4,
\end{equation}
and vary with respect to $\sigma$. One finds
\begin{equation}
\partial_t^2 \sigma = \frac{2}{3} \vec \nabla^2 \sigma - 2 g^2 \sigma^3.
\label{eq:EOMsigma}
\end{equation}
For $\sigma$ obeying the equation of motion \eqref{eq:EOMsigma} the energy density \eqref{eq:nergypressuresigma} is a conserved quantity. 

\begin{figure}
\begin{center}
\includegraphics[width=0.5\textwidth]{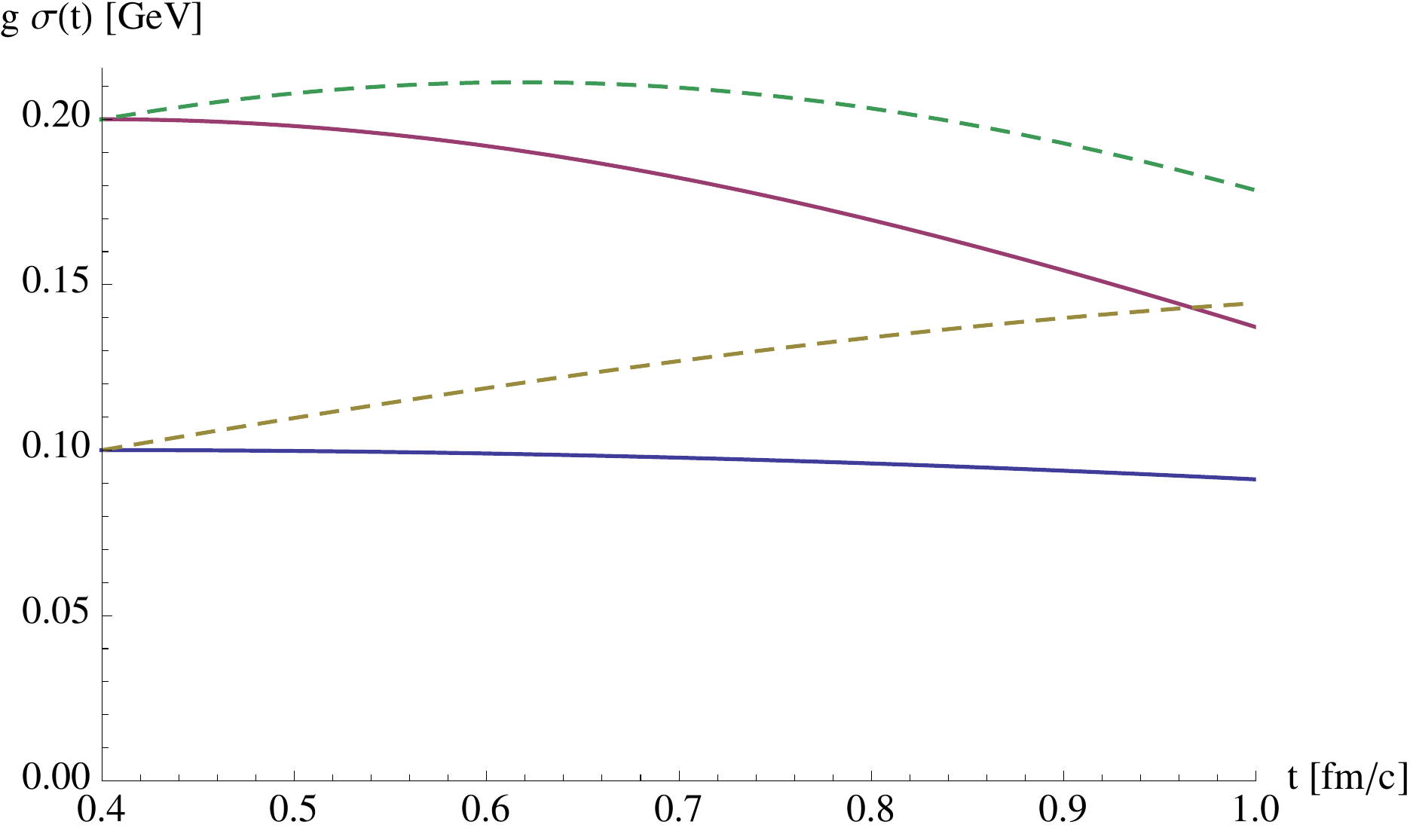}
\caption{Time evolution of the isotropic condensate $\sigma(t)$ in a homogeneous situation with different initial values $\sigma(t_\text{in})$ and $\dot\sigma(t_\text{in})$.}
\label{fig6}
\end{center}
\end{figure}

For vanishing spatial gradients eq.\ \eqref{eq:EOMsigma} becomes the equation for an anharmonic oscillator,
\begin{equation}
\frac{d^2}{d t^2} (g \sigma) = - 2 (g \sigma)^3,
\label{eq:eomsigma}
\end{equation}
that can be solved in terms of Jacobi elliptic functions. Eq.\ \eqref{eq:eomsigma} is of the same structure as the time evolution equation for a homogeneous $SU(2)$ gauge field with a specific orientation in position and color space as investigated in ref.\ \cite{Berges:2011sb} where one can also find a detailed discussion of its solution. In Fig.\ \ref{fig6} we plot the solution $\sigma(t)$ for different values of $\sigma(t_\text{in})$ and $\dot \sigma(t_\text{in})$. Note that larger initial values $\sigma(t_\text{in})$ lead to a faster oscillating behavior. Small values $g\sigma(t_\text{in})\ll 1\,\text{GeV}$ are practically constant over the time range of  $1 \, \text{fm/c}$. 
We also show two curves with positive $\dot \sigma(t_\text{in})$ as suggested by Fig.\ \ref{fig7}. To be explicit we have taken $t_\text{in}=0.4 \, \text{fm/c}$, where the $\sigma$-condensate starts to dominate in fig.\ \ref{fig7}. In view of this qualitative behavior ``survival times'' of the $\sigma$-condensate of $1\, \text{fm/c}$ or larger seem quite natural.

\begin{figure}
\begin{center}
\includegraphics[width=0.5\textwidth]{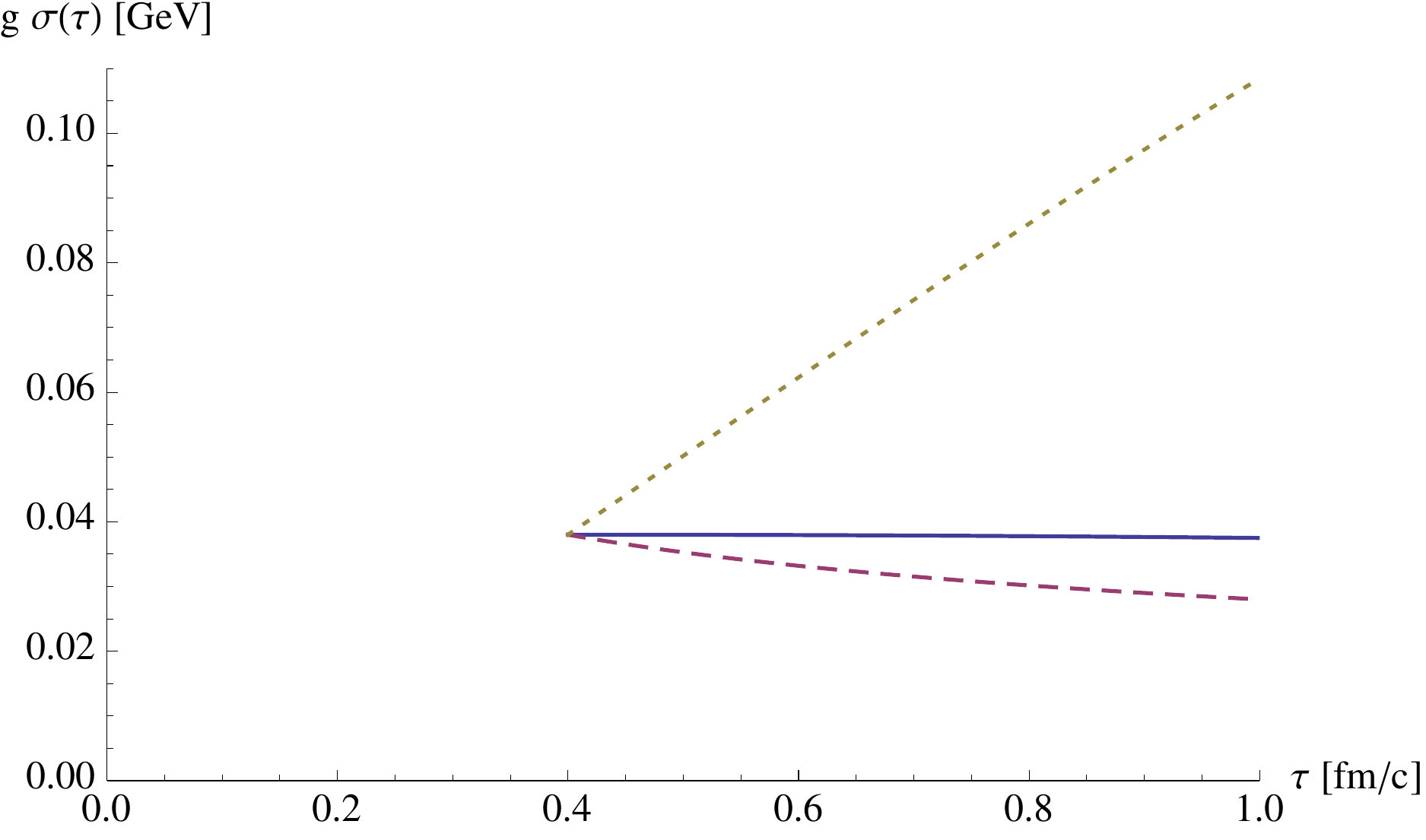}
\caption{Time evolution of condensate $\sigma(t)$ in a homogeneous situation, initialized at $t_\text{in}=0.4 \, \text{fm/c}$ by the value of $\sigma(t_\text{in})$ taken from the solution in Fig.\ \ref{fig7}  (solid line). We also show the case where both $\sigma(t_\text{in})$ and $\dot\sigma(t_\text{in})$ are taken over from the solution in Fig.\ \ref{fig7} (dotted line).
These solutions do not take a longitudinal expansion into account. We compare this to a power-law decay, $\sigma\sim t^{-1/3}$ (dashed line) as obtained for a Bjorken expansion with isotropic pressure and under the assumption that the energy density is dominated by the potential energy of the condensate, $\epsilon=6 g^2 \sigma^4$. }
\label{fig11}
\end{center}
\end{figure}

In a real heavy ion collision the time evolution equations will be modified by different effects. At early times the most prominent one might be the longitudinal expansion. The question arises wether the oscillating behavior according to eq.\ \eqref{eq:eomsigma} is more or less important than the dilution effect caused by the longitudinal expansion.
As a simple guideline we compare in Fig.\ \ref{fig11} two very simple scenarios. For the first the time evolution of the condensate neglects Bjorken expansion. It is initialized at $\tau=0.4\,\text{fm/c}$ with the values of $\sigma(t_\text{in})$  taken from the solution in Fig.\ \ref{fig7} (solid line). We also show the solution obtained when both $\sigma(t_\text{in})$ and $\dot \sigma(t_\text{in})$ are taken over from Fig.\ \ref{fig7} (dotted line). The second scenario is a power law decay, $\sigma\sim t^{-1/3}$,  as one gets it for a Bjorken expansion $\partial_t \epsilon + \frac{\epsilon+p}{t}=0$ with isotropic pressure $p=\epsilon/3$ under the assumption that the energy density is dominated by the potential energy of the condensate, $\epsilon=6 g^2 \sigma^4$ (dashed line). This comparison seems to suggest that the dilution effects by the longitudinal expansion are comparable to the time dependence due to the non-linear term.
We conclude that the dilution effect is not strong enough to reduce the ``survival time'' below 1 fm/c.

We next proceed to a similar analysis in case of a reduced cylindrical symmetry. For the homogeneous condensates $\sigma$, $\tilde \gamma^A$ and $\tilde \gamma^B$ one obtains the equations of motion
\begin{equation}
\begin{split}
\frac{d^2}{dt^2} \sigma = & g^2 \left( \tilde \gamma^B - \sigma \right) \left( 9 (\tilde \gamma^A)^2 + (\tilde \gamma^B + \sigma)^2 \right)\\
& - g^2 \left( \tilde \gamma^B + \sigma \right) \left( 3 (\tilde \gamma^A)^2 + 2 (\tilde \gamma^B)^2 + (\tilde \gamma^B - \sigma)^2 \right),\\
\frac{d^2}{dt^2} \tilde \gamma^A = & - \frac{1}{2} g^2 \tilde \gamma^A \left( 9 (\tilde \gamma^A)^2 + (\tilde \gamma^B + \sigma)^2 \right)\\
& -\frac{3}{2} g^2 \tilde \gamma^A \left( 3 (\tilde \gamma^A)^2 + 2 (\tilde \gamma^B)^2 + (\tilde \gamma^B - \sigma)^2 \right),\\
\frac{d^2}{dt^2} \tilde \gamma^B = & - \frac{1}{2} g^2 \left(3 \tilde \gamma^B-\sigma \right) \left( 9 (\tilde \gamma^A)^2 + (\tilde \gamma^B + \sigma)^2 \right)\\
& -\frac{1}{2} g^2 \left( \tilde \gamma^B+\sigma \right) \left( 3 (\tilde \gamma^A)^2 + 2 (\tilde \gamma^B)^2 + (\tilde \gamma^B - \sigma)^2 \right).
\end{split}
\label{eq:timeevolutioncondensates}
\end{equation}
For $\tilde \gamma^A=\tilde \gamma^B=0$ the first equation is equivalent to \eqref{eq:eomsigma} while the right hand side of the second and third equation vanishes as expected from symmetry considerations.

 In Fig.\ \ref{fig7} we show the time evolution for a situation where the isotropic condensate vanishes initially, $\sigma(0)=0$, but is generated by the time evolution. In this example we have chosen the initial values of $\tilde \gamma^A(0)$ and $\tilde \gamma^B(0)$ such that that the initial longitudinal pressure $p_l$ vanishes and therefore $\epsilon=2p_{tr}$. 
The time evolution of the transverse pressure $p_{tr}(t)$, the longitudinal pressure $p_l(t)$ and the conserved energy density $\epsilon$ are shown in Fig.\ \ref{fig8}. It is interesting to see that $p_l$ and $p_{tr}$ approach each other rather quickly and approximate isotropization is reached at times of about $0.2\, \text{fm/c}$. In a realistic expansion after a heavy ion collision additional effects become important. The longitudinal and transverse expansions have the tendency to reduce all condensates by dilution. One expects that this results in an effective damping of the oscillations. Furthermore, other non-symmetric and non-homogeneous modes will by excited. A superposition of such modes has again a tendency to reduce any oscillatory behavior. Also the production of particles with larger momenta will damp the color field condensate. Nevertheless, the basic mechanism that an initial field configuration triggers the onset of a rotation symmetric configuration by effective source terms in the field equation for $\sigma$ remains unaffected. The general tendency that the dynamics of field condensates drives the energy momentum tensor towards an isotropic situation should be well captured by our model. On the other side we do not see a strong reason to deviate from isotropy again after it has been reached.

\begin{figure}
\begin{center}
\includegraphics[width=0.5\textwidth]{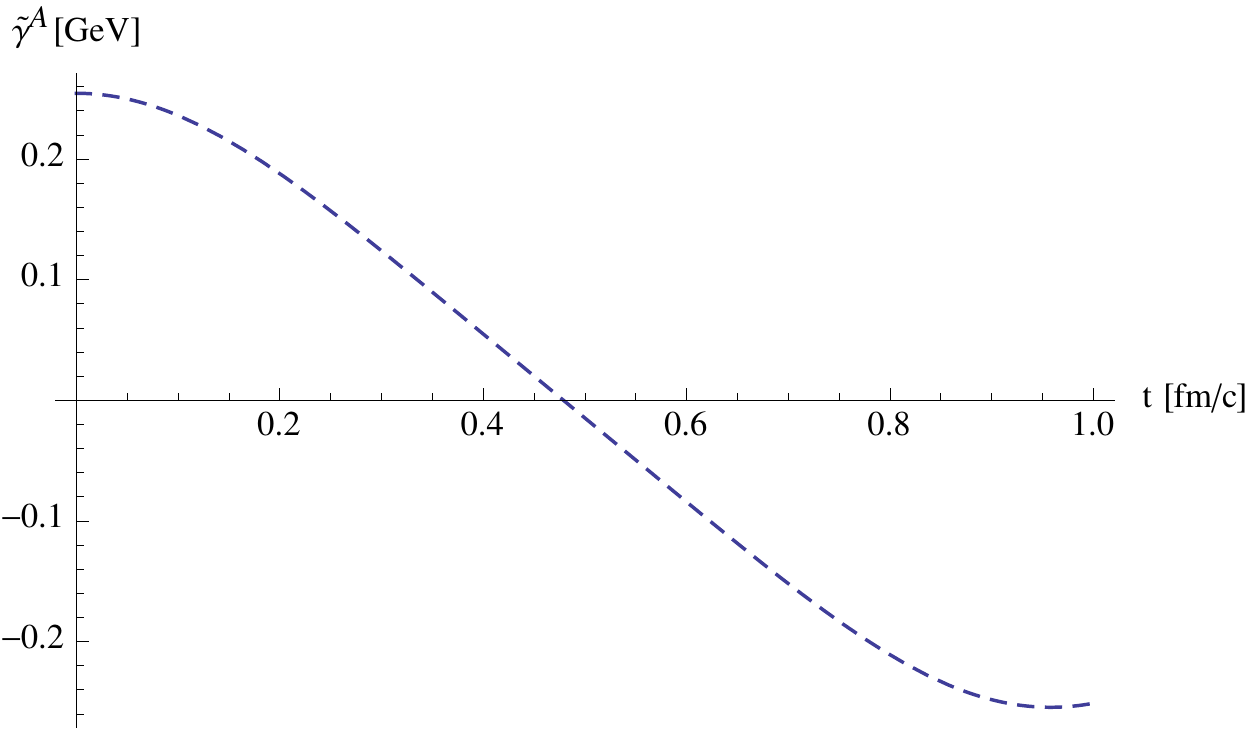}
\caption{Time evolution of the $\mathsf{CP}$-even cylindrical condensate $\gamma^A(t)$ (dashed line) for an initialization where the isotropic condensate $\sigma$ and the $\mathsf{CP}$-odd cylindrical condensate $\tilde \gamma^B$ (and their time derivatives) vanish initially and therefore at all times.}
\label{fig9}
\end{center}
\end{figure}

\begin{figure}
\begin{center}
\includegraphics[width=0.5\textwidth]{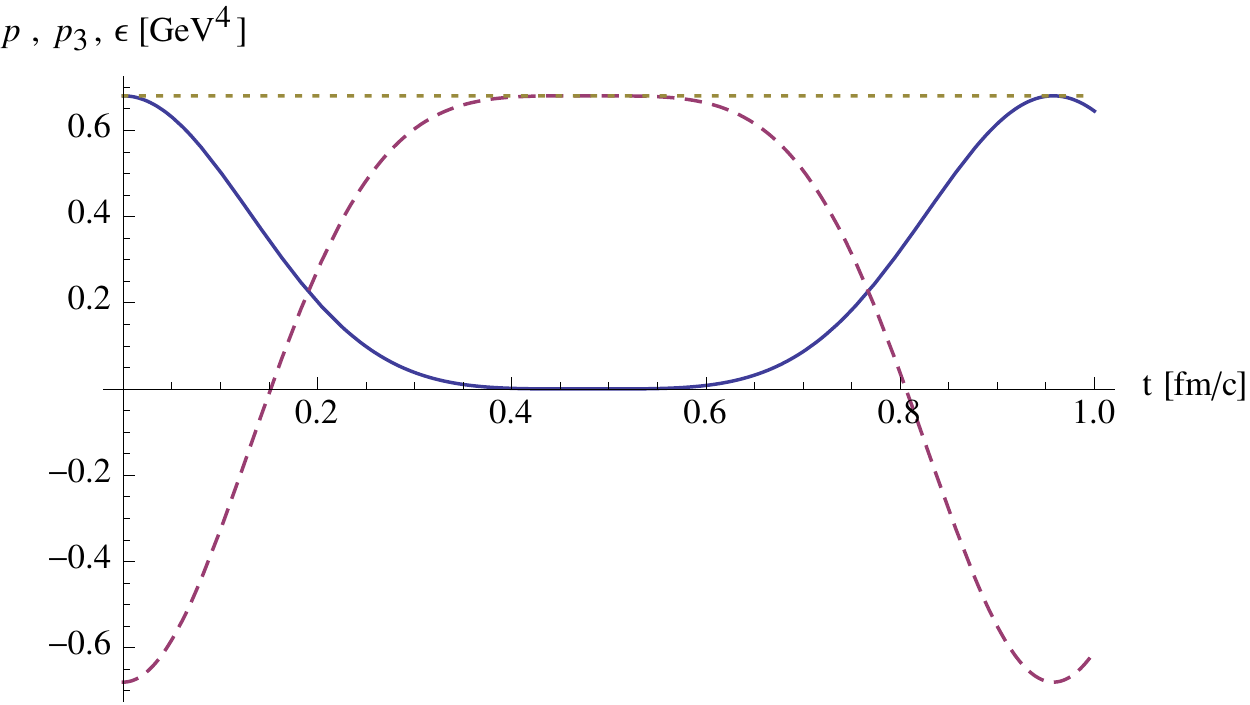}
\caption{Time evolution of the transverse pressure $p_{tr}(t)$ (solid line),  longitudinal pressure $p_l(t)$ (dashed line) and energy density $\epsilon(t)$ (dotted line) corresponding to the time dependent cylindrical condensate $\tilde \gamma^A$ as shown in Fig.\ \ref{fig9}.}
\label{fig10}
\end{center}
\end{figure}

In Fig.\ \ref{fig9} we plot another interesting solution of the evolution equations \eqref{eq:timeevolutioncondensates}. Here we have chosen the initial isotropic condensate and it's time derivative to vanish, $\sigma(0)=\dot \sigma(0)=0$ and similar for one of the cylindrical condensates, $\tilde \gamma^B(0) = \dot{\tilde \gamma}^{B}(0)=0$. Together with $\dot{\tilde \gamma}^{A\prime}(0)=0$ this implies that the initial energy momentum tensor is of the form $\text{diag}(\epsilon,\epsilon,\epsilon,-\epsilon)$.
From the evolution equations \eqref{eq:timeevolutioncondensates} it follows that $\sigma$ and $\tilde \gamma^B$ are not generated in this case.
The time dependence of the transverse and longitudinal pressure in this case are shown in Fig. \ref{fig10}. Interestingly, the solution shows a periodic behavior where first the longitudinal pressure quickly grows and becomes positive while the transverse pressure decreases. An isotropic situation is again reached at about $t=0.2\, \text{fm/c}$. The longitudinal pressure keeps growing, however and at some point one reaches a situation where $\epsilon=p_l$ and where the transverse pressure $p_{tr}$ vanishes. 
Again we believe that this behavior will be modified in a realistic collision but the initial approach towards an isotropic energy-momentum tensor may reflect qualitatively a realistic situation. 

The most important lesson that we can infer from these qualitative investigations is very simple: the characteristic time scale associated to isotropization is very short in the presence of substantial color field condensates. Collective effects can lead to fast isotropization! We observe that an almost isotropic pressure may be reached only 0.2 fm/c after the collision -- much faster than by the scattering of individual gluons and quarks.
\section{Excitations}
\label{sec:Excitations}

In the following two sections we investigate small deviations from the isotropic color field condensate. The stability analysis of these fluctuations serves two purposes. First, the behavior of fluctuations with small momenta tells us about the stability of the homogeneous isotropic configuration with respect to neighboring homogeneous, non-isotropic configurations. The relevant quantity to study is the dispersion relation for a given fluctuation
\begin{equation}
\omega^2 = \varepsilon^2(\vec p).
\label{eq:F1}
\end{equation}
Positive $\varepsilon^2(\vec p=0)$ ensures stability with respect to homogeneous fluctuations. In this case effective source terms for the fluctuations which arise, for example, from gradients generated due to the expansion, induce only small anisotropies of the field configuration. In the presence of effective damping (due to effects beyond our approximation such as the expansion of the fireball) initial anisotropies tend to be reduced. Typical classical solutions are then damped oscillations around the isotropic color field.

In contrast, for negative $\omega_0^2=\varepsilon^2(\vec p=0)$ (i.\ e.\ imaginary $\omega_0$) the isotropic solution is unstable. Anisotropies in the color field will appear within a typical time given by $1/|\omega_0|$. For too large negative $\omega_0^2$ the isotropic condensate $\sigma$ would not be a reasonable candidate for a good description of the qualitative features of isotropization. We find that $\omega_0^2$ is positive or zero for all fluctuations if we assume eq.\ \eqref{eq:YMLagrangian} as an approximation to the relevant time dependent effective action. This strengthens the point that the condensate \eqref{eq:singletconfigurationexplicit} may indeed play an important role for isotropization. A similar study for the alternative configuration \eqref{eq:2.2} is not yet available.

A second issue concerns a possible range of non-zero momenta $\vec p$ for which $\varepsilon^2(\vec p)$ may turn negative for some modes. This type of instability has a different interpretation if momenta are close to ``thermal momenta'' $p_\text{Temp}$ and the negative values of $\varepsilon^2(\vec p)$ concern a whole momentum range. Rather than indicating the generation of a particular coherent inhomogeneous field, such an instability should be interpreted as a sign of exponential particle production, with typical momenta of particles in the range where $\varepsilon^2(\vec p)$ is negative. This indicates a (partial) decay of the color field condensate into particles. There is no reason why local isotropy should be broken by particle production. One simply expects that the isotropic energy momentum tensor of the color field is transformed to an isotropic energy momentum tensor describing particles plus a remnant color field. This isotropic particle production is an important ingredient for the final emergence of a quark gluon plasma that shares many properties of a thermal equilibrium state.

The overall picture depends on the typical time scale $t_{pp}$ of particle production. In a rough approximation it may be given by $t_{pp}=1/|\varepsilon(p_\text{min})|$ if $\varepsilon(\vec p)$ has a minimum with negative values. For large $t_{pp}$, say $t_{pp}> 1\, \text{fm/c}$, the prethermalization of the fireball may occur in two stages. The first stage produces a color field with almost isotropic energy momentum tensor and achieves approximate isotropization. In the second stage the isotropic gluon field condensate decays into a locally isotropic quark gluon plasma. For $t_{pp}$ substantially smaller than $1\,\text{fm/c}$ the distinction between the two stages would be washed out.

In Fig.\ \ref{fig2} we display the dispersion relation for a typical fluctuation. It shows stability for small momenta, with $\omega_0^2=\varepsilon^2(\vec p=0)= 2 g^2 \sigma^2$. For one of the two modes the function $\varepsilon^2(\vec p)$ gets negative, however, in a range of higher momenta, with minimum for $|p_\text{min}|=2 \sqrt{g^2 \sigma^2}$. The value at the minimum is $\varepsilon^2(p_\text{min}) = -2 g^2 \sigma^2$. For a typical value $g\sigma \approx 0.1\,\text{GeV}$ this indicates a time scale for particle production $t_{pp}$ somewhat above 1 fm/c.

Even in the case of the effective action \eqref{eq:YMLagrangian} the computation of the dispersion relations for fluctuations around the isotropic color field is rather involved. While straightforward in principle, it involves a rather complex diagonalization problem that we describe in this and the following sections. Dispersion relations of the simple form \eqref{eq:F1} can only be obtained after decomposing the fluctuations into modes according to their transformation properties with respect to their symmetries, eliminating gauge modes, implementing constraints, and finally diagonalizing the stability matrix. We also recall at this place that the effective action \eqref{eq:YMLagrangian} is at best an approximation for early stages of the prethermalization process. Once the system gets closer to thermal equilibrium the effective action for a thermal equilibrium system will be a more appropriate approximation. This may change the qualitative features. It is conceivable that a color field condensate with a temperature dependent ``thermal value'' $\sigma_\text{Temp}$ is present in equilibrium and stable. 

\subsection{Modified rotation symmetry}
\label{sec:ModifiedRotationSymmetry}

In this paper we use a modified form of rotations under which the gluon field does not transform as a vector $\bold {A}_j$ and a scalar $\bold{A}_0$. Different components of the gluon fields transform as different tensors, with up to three indices. This is due to the gauge transformation that accompanies rotations. We start our discussion by recalling how infinitesimal rotations act on a scalar gauge singlet $\phi$,
\begin{equation}
\phi(x_0,\vec x) \to \phi(x_0,\vec x^\prime)
\end{equation}
with
\begin{equation}
x^\prime_i = \left(\delta_{ij}-i \alpha_l (\hat t^l)_{ij} \right) x_j,
\end{equation}
and $\bold{\hat t}^l$ the standard generators of the SO(3) Lie algebra,
\begin{equation}
(\hat t^l)_{ij} = -i \epsilon_{lij}.
\end{equation}
For gauge singlet vector fields this is generalized to the standard transformation of a vector field.
In contrast, for the gluon field $\bold{A}_\mu=A_\mu^a \bold{t}^a$ space rotations are accompanied by gauge rotations such that $\bold{A}_\mu$ transforms as
\begin{equation}
\begin{split}
\delta (A_0)_{pq}(x_0,\vec x) = & i \alpha_l \left[ (\hat t^l)_{pm} \delta_{qn} +\delta_{pm} (\hat t^l)_{qn} \right]   (A_0)_{mn}(x_0,\vec x^\prime),\\
\delta (A_j)_{pq}(x_0,\vec x)  = & i \alpha_l \left[(\hat t^l)_{ji}\delta_{pm} \delta_{qn}+\delta_{ji}  (\hat t^l)_{pm} \delta_{qn}+\delta_{ji} \delta_{pm} (\hat t^l)_{qn} \right] (A_i)_{mn}(x_0,\vec x^\prime).
\end{split}
\label{eq:transformationA}
\end{equation}
The indices $p,q$ and $m,n$ are the color indices with possible values $1,2,3$. Only the first term $\sim (\hat t^l)_{ji}$ in the second equation \eqref{eq:transformationA} is related to standard space rotations, while all other infinitesimal variations arise from the special global gauge transformations,
\begin{equation}
\delta \bold{A}_\mu = i [\bold{\beta}, \bold{A}_\mu ]
\end{equation}
with non-zero elements of the hermitian matrix $\bold{\beta}$ given by $\beta_{12} = -\beta_{21} = -i \alpha_3$, $\beta_{23}=-\beta_{32} = -i \alpha_1$, $\beta_{31}=-\beta_{13}=-i \alpha_2$.
Being in the adjoint representation of $SU(3)$, the gauge field is a hermitian and traceless matrix, 
\begin{equation}
\bold{A}_\mu=\bold{A}_\mu^\dagger,\quad \text{tr } \bold{A}_\mu=0.
\end{equation}

\subsection{Decomposition of gauge field}
\label{sec:Decompositionofgaugefield}
Equation \eqref{eq:transformationA} states that the spatial components of the gluon field $\bold{A}_j$ transform as a (reducible) tensor of rank three under the modified rotation symmetry. It can be decomposed as
\begin{equation}
(A_j)_{mn}=\kappa_{jmn} +\gamma^A_{mk}\epsilon_{kjn} + \gamma^A_{nk} \epsilon_{kjm}+ i\, \gamma^B_{jk} \epsilon_{kmn} + (\beta^A_m+i\,\beta^B_m) \,\delta_{jn} + (\beta^A_n-i\,\beta^B_n)\, \delta_{jm} - \frac{2}{3}\beta^A_j \delta_{mn}  + i\, \sigma\, \epsilon_{jmn}.
\label{eq:decompositionAspatial}
\end{equation}
The real field $\kappa_{jmn}$ is a completely symmetric tensor of rank three which is traceless with respect to all possible contractions. Similarly $\gamma^A_{jk}$ and $\gamma^B_{jk}$ are real, symmetric and traceless. The vector-type fields $\beta^A_m$ and $\beta^B_m$ are real and $\sigma$ is a real scalar field. We have thereby decomposed the $3\times 8$ spatial components of the gauge field according to
\begin{equation}
\bold{24} = \bold{7} + 2\times \bold{5} + 2\times\bold{3} + \bold{1}.
\end{equation}

Now we turn to the temporal component $\bold{A}_0$. It transforms as a tensor of rank two which can be decomposed into
\begin{equation}
(A_0)_{mn}=\gamma^{C}_{mn} + i\, \beta^{C}_l \epsilon_{lmn},
\label{eq:decompositionAtemporal}
\end{equation}
where $\gamma^{C}_{mn}$ is real, symmetric and traceless and $\beta^{C}_l$ is real. Thus, the temporal component of the gauge field is decomposed according to 
\begin{equation}
\bold{8}=\bold{5}+\bold{3},
\end{equation}
and the the complete field $\bold{A}_\mu$ is reduced as
\begin{equation}
\bold{32} = \bold{7} + 3\times \bold{5} + 3\times \bold{3} + \bold{1}.
\end{equation}

It is clear that in a situation with three dimensional rotation invariance only the singlet field $\sigma$ can be non-zero. A cylindrical symmetry, i.\ e.\ rotation invariance in the $x_1$-$x_2$-plane together with invariance under rotations of $180$ degree around the $x_1$ axis, $(x_1,x_2,x_3) \to (x_1,-x_2,-x_3)$ and around the $x_2$ axis, $(x_1,x_2,x_3) \to (-x_1, x_2,-x_3)$, allows also non-zero values of the symmetric and traceless tensors of the form
\begin{equation}
\gamma^A_{mn} = 
\begin{pmatrix}
\tilde \gamma^A, & 0, & 0 \\
0, & \tilde \gamma^A, & 0 \\
0, & 0, & -2 \tilde \gamma^A
\end{pmatrix}, \quad \quad
\gamma^B_{mn} = 
\begin{pmatrix}
\tilde \gamma^B, & 0, & 0 \\
0, & \tilde \gamma^B, & 0 \\
0, & 0, & -2 \tilde \gamma^B
\end{pmatrix},
\end{equation}
and similar for $\gamma^C_{mn}$.

The decomposition in \eqref{eq:decompositionAspatial} and \eqref{eq:decompositionAtemporal} of $\bold{A}_j$ and $\bold{A}_0$ into tensor, vector and scalars under the modified rotation symmetry is not yet sufficient. The dynamical evolution can mix different terms, for example the scalar can mix with $\partial_l \beta_l^i$ etc. To facilitate the analysis of the evolution equations we go therefore a step further and write
\begin{equation}
\beta^i_l = \partial_l  \beta^i+ \hat \beta^i_l, \quad \quad i=A,\ldots,C,
\label{eq:decompBeta}
\end{equation}
where $\beta^i$ are real scalars and $\hat \beta^i_l$ are real, divergence-less vectors, $\partial_l \hat \beta_l^i=0$. It is clear that the constant part of $\beta^i$ does not have a physical meaning, or, in other words, there is a global shift symmetry $\beta^i\to\beta^i+c^i$. With respect to the modified rotation symmetry, equation \eqref{eq:decompBeta} corresponds to the splitting into $2+1$ degrees of freedom.

Similarly we decompose the symmetric and traceless rank-two tensors
\begin{equation}
\gamma_{mn}^i =  \hat \gamma^i_{mn} + \partial_m \hat\gamma_n^i + \partial_n \hat\gamma^i_m + (\partial_m \partial_n - \frac{1}{3}\delta_{mn} \partial_j^2)  \gamma^i, \quad \quad i=A,\ldots,C,
\label{eq:decompGamma}
\end{equation}
with $\hat \gamma^i_{mn}$ real, traceless and divergence-less tensors, $\partial_m \hat \gamma^i_{mn}=0$, $\hat \gamma^i_l$ real, divergence-less vectors, $\partial_l \hat \gamma^i_l=0$ and $\gamma^i$ a real scalar. We use the abbreviation $\partial_j^2 = \sum_j \partial_j \partial_j$. Again there are shift symmetries $\hat \gamma^i_l\to \hat \gamma^i_l + c^i_l$ where $c^i_l$ is constant and $\gamma^i\to\gamma^i+c^i$  where $c^i$ is a linear function of space. Eq.\ \eqref{eq:decompGamma} corresponds to the splitting of degrees of freedom according to $5=2+2+1$.

Finally, the symmetric and traceless rank-three tensor can be written
\begin{equation}
\begin{split}
\kappa_{jmn} = & \hat \kappa_{jmn} + \partial_j \hat \kappa_{mn} + \partial_m \hat \kappa_{jn} + \partial_n \hat \kappa_{jm}\\ 
& + (\partial_j \partial_m - \frac{1}{3} \delta_{jm} \partial_k^2) \hat \kappa_n
+ (\partial_m \partial_n - \frac{1}{3} \delta_{mn} \partial_k^2) \hat \kappa_j
+ (\partial_n \partial_j - \frac{1}{3} \delta_{nj} \partial_k^2) \hat \kappa_m\\
& + \left[\partial_j\partial_m \partial_n - \frac{1}{3} \partial_k^2 ( \delta_{mn} \partial_j + \delta_{jn} \partial_m +  \delta_{jm} \partial_n) \right]  \kappa,
\end{split}
\label{eq:decompKappa}
\end{equation}
where $\hat \kappa_{jmn}$ is real, symmetric, traceless and divergence-less with respect to all indices, $\hat \kappa_{mn}$ is real, symmetric, traceless and divergence-less with shift symmetry $\hat \kappa_{mn} \to \hat \kappa_{mn}+c_{mn}$, $\hat \kappa_m$ is a real, divergence-less vector with shift symmetry $\hat \kappa_m \to \hat \kappa_m + c_m$ where $c_m$ is a linear function of space and similarly $\kappa$ is a real scalar with shift symmetry $\kappa \to \kappa + c$ where $c$ is now a maximally quadratic function of space. This splitting corresponds to $7=2+2+2+1$ independent degrees of freedom.

Rotation symmetry implies that scalars, divergence-less vectors, traceless second and divergence-less second and third rank tensors cannot mix. The (linearized) field equations become block-diagonal in these components. On the other side, mixing between the different fields of a given type is possible and expected. For example, the 8 scalars $\kappa$, $\gamma^i$, $\beta^i$ and $\sigma$ are allowed to mix. Some of these fields will be identified below as gauge degrees of freedom, but the remaining propagator for the physical fields will have to be diagonalized.

In addition to the continuous rotation symmetry there are discrete symmetries that can be used to classify the different representations in \eqref{eq:decompositionAspatial} and \eqref{eq:decompositionAtemporal}. This will further split the blocks into smaller sub-blocks. For example charge conjugation even end odd fields cannot mix. Indeed, it is easy to see that the Yang-Mills Lagrangian is invariant under ``color charge conjugation'' $\mathsf{C}$
\begin{equation}
\bold{A}_\mu \to - \bold{A}_\mu^*.
\end{equation}
The fields $\kappa_{jmn}$, $\gamma^A_{mn}$, $\beta^A_m$ and $\gamma^C_{mn}$ are odd with respect to this transformation while the fields $\gamma^B_{mn}$, $\beta^B_m$, $\sigma$ and $\beta^C_m$ are even.

Another discrete symmetry is parity $\mathsf{P}$, $(x_0,x_j) \to (x_0,-x_j)$. With respect to space reflections the gluon field transforms as a vector,
\begin{equation}
(\bold{A}_0, \bold{A}_j) \to (\bold{A}_0, -\bold{A}_j),
\label{eq:parityTrans}
\end{equation}
such that the fields $\kappa_{jmn}$, $\gamma^A_{mn}$, $\gamma^B_{mn}$, $\beta^A_m$, $\beta^B_m$ and $\sigma$ are odd while $\gamma^C_{mn}$ and $\beta^C_{m}$ are even. A non-vanishing isotropic condensate $\sigma$ or cylindrical condensate $\tilde \gamma^A$, $\tilde \gamma^B$ does break parity. One can show that time inversion $\mathsf{T}$ is broken in a similar way such that the combined symmetry $\mathsf{CPT}$ is conserved.

One may wonder whether space parity reflections can be combined with a suitable discrete gauge transformation such that the color field condensate \eqref{eq:singletconfigurationexplicit} remains invariant. This would require the existence of a unitary matrix $U_\mathsf{P}$, $U_\mathsf{P}^\dagger U_\mathsf{P}=1$, with the property
\begin{equation}
U_\mathsf{P} \, \bold{A}^{(\sigma)}_j \, U_\mathsf{P}^\dagger = - \bold{A}^{(\sigma)}_j.
\label{eq:G2}
\end{equation}
Here $\bold{A}^{(\sigma)}$ is the color field according to eq.\ \eqref{eq:singletconfigurationexplicit}, and \eqref{eq:G2} has to hold for all space like components $\bold{A}_j$. If a matrix obeying \eqref{eq:G2} can be found we can define a combined parity from the transformations \eqref{eq:G2} and \eqref{eq:parityTrans} under which $\sigma$ is invariant. For gauge singlets this modified parity would correspond to standard parity transformations. A matrix $U_\mathsf{P}$ obeying eq.\ \eqref{eq:G2} would have to anti-commute with the three Gell-Mann matrices $\lambda_2$, $\lambda_5$ and $\lambda_7$. No such matrix exists, however, and we conclude that that the isotropic color field \eqref{eq:singletconfigurationexplicit} indeed breaks parity and time reversal. While this is possible for the non-equilibrium situation of the expanding fireball, it seems less likely that such a parity violating condensate persists in thermal equilibrium.

The situation is actually similar for the alternative isotropic gauge field \eqref{eq:2.2}. Again, the singlet $\sigma$ is odd under space-parity transformations $\mathsf{P}$. No unitary matrix exists which anti-commutes simultaneous with $\lambda_1$, $\lambda_2$ and $\lambda_3$, such that no modified parity can be defined with $\sigma$ even. With respect to color charge conjugation $\mathsf{C}$ the components $\bold{A}^{(\sigma)}_1$ and $\bold{A}^{(\sigma)}_3$ change sign while $\bold{A}^{(\sigma)}_2$ is invariant. One can find a unitary matrix such that $U_\mathsf{C} \bold{A}^{(\sigma)}_{1,3} U_\mathsf{C}^\dagger = - \bold{A}^{(\sigma)}_{1,3}$ and $U_\mathsf{C} \bold{A}^{(\sigma)}_{2} U_\mathsf{C}^\dagger =  \bold{A}^{(\sigma)}_{2}$. This is realized by
\begin{equation}
U_\mathsf{C}=\begin{pmatrix} 0, & -i, & 0 \\ i, & 0, & 0 \\ 0, & 0, & 1
\end{pmatrix}.
\end{equation}
Thus for the ansatz \eqref{eq:2.2} $\sigma$ remains invariant if color charge conjugation is combined with this particular gauge transformation.
\section{Stability analysis}
\label{sec:StabilityAnalysis}

For a stability analysis we have to identify the physical modes that correspond to excitations on the on the isotropic background \eqref{eq:singletconfigurationexplicit}. Conceptually, one linearizes the Yang-Mills field equations
\begin{equation}
[\bold{D}_\nu, \bold{F}^{\mu\nu}] = 0, \quad \quad \bold{D}_\nu = \partial_\nu - i g \bold{A}_\nu,
\label{eq:F1}
\end{equation}
in small deviations of the gauge field from the background. (The background obeys itself eq.\ \eqref{eq:F1}.) We concentrate on the case where the time dependence of $\sigma$ is small enough to be neglected. In other words, we linearize around a homogeneous and static background \eqref{eq:singletconfigurationexplicit}. This approximation assumes also that a possible parametric resonance due to the oscillating behavior of the condensate, which was shown to lead to a subdominant instability band in ref.\ \cite{Berges:2011sb}, has only a weak effect. To investigate the implications of this parametric resonance in our formalism one would have to solve the time evolution equations for the fluctuations around an oscillating background. For a constant background one can use gauge fixing in order to eliminate the pure gauge modes. We will use the Weyl gauge $\bold{A}_0=0$.

In practice, we proceed somewhat differently. We first evaluate the action up to terms quadratic in the non-singlet fluctuations, inserting $\bold{A}_0=0$. Rotation symmetry forbids that there are terms linear in the non-singlet fields, and we may call the result the gauge fixed quadratic action. Any solution of the linearized eq.\ \eqref{eq:F1} has to obey the equations that follow from the variation of this quadratic action. The linearized field equations \eqref{eq:F1} also contain an equation which arises from the variation of the action \eqref{eq:YMLagrangian} with respect to $\bold{A}_0$, corresponding to $\mu=0$ in eq.\ \eqref{eq:F1}. This equation is not contained in the variation of the gauge fixed quadratic action, since $\bold{A}_0$ has been set before the variation. The $\mu=0$ component of \eqref{eq:F1} has therefore to be imposed in addition. It will act as an additional constraint for the solutions of the field equations derived from the gauge fixed quadratic action. Once this constraint is imposed we can diagonalize the quadratic action and extract the dispersion relations for the physical modes.

\subsection{Gauge fixing}

Infinitesimal SU(3) gauge transformations act on the field $\bold{A}_\mu=A_\mu^a \bold{t}^a$ according to
\begin{equation}
\delta A_\mu^a =  \frac{1}{g} \partial_\mu \alpha^a + f^{abc} A_\mu^a \alpha^c = \frac{1}{g} D_\mu \alpha^a
\end{equation}
or, in a representation as in \eqref{eq:transformationA} with $\alpha_{mn} = \alpha^a (t^a)_{mn}$ a hermitean and traceless $2\times 2$ matrix,
\begin{equation}
\begin{split}
\delta (A_\mu)_{mn} = \frac{1}{g} \partial_\mu \alpha_{mn} - i (A_\mu)_{mp} \alpha_{pn} + i \alpha_{mp} (A_\mu)_{pn}.
\end{split}
\end{equation}
We now write the matrix $\alpha_{mn}$ as
\begin{equation}
\alpha_{mn} = \chi_{mn} + i \,\zeta_l \, \epsilon_{lmn}.
\end{equation}
Here $\chi_{mn}$ is a real, symmetric and traceless matrix and $\zeta_l$ is a real vector. Different gauges can be fixed by making appropriate choices of $\chi_{mn}$ and $\zeta_l$. 

Gauge fixing conditions that might be particularly useful in the present context are Weyl gauge $(A_0)_{mn}=0$, or Coulomb gauge, $D_j(A_j)_{mn}=0$. In terms of the decomposition \eqref{eq:decompositionAspatial}, \eqref{eq:decompositionAtemporal} the Weyl gauge corresponds to $\gamma^C_{mn} = \beta^C_l=0$. We will adopt Weyl gauge and therefore eliminate the gauge degrees of freedom $\gamma^C_{mn}$ and $\beta^C_l$. One observes that the Weyl gauge is not complete in the sense that it fixes only the time dependence of $\alpha_{mn}$.  One can in addition choose a time-independent $\alpha_{mn}$ such that also $D_j (A_j)_{mn} = 0$ is obeyed for a given time $t$. On the level of the linearized field equations this property is conserved due to constraint equations, see appendix \ref{app:ConstraintEquations}.

\subsection{Decomposition of the Lagrangian in Weyl gauge}
Next we discuss the form of the Lagrangian in terms of those fields that form irreducible representations of the modified rotation symmetry. The gluon contribution to the Lagrangian in the presence of an external source is
\begin{equation}
{\cal L} = - \frac{1}{2} \text{tr} 
\,\bold{F}^{\mu\nu} \bold{F}_{\mu\nu} + 2\, \text{tr}\,  {\bold A}_\mu {\bold J}^\mu = \text{tr} ( \bold{E}_j \bold{E}_j - \bold{B}_j \bold{B}_j ) + 2\, \text{tr}\,  {\bold A}_\mu {\bold J}^\mu.
\end{equation}
Here we have added sources $\bold{J}^\mu$ that can also be decomposed in term of irreducible representations, see appendix \ref{app:DecompositionSourceTerms}.
The field strength tensor
\begin{equation}
\bold{F}_{\mu\nu} = \partial_\mu \bold{A}_\nu - \partial_\nu \bold{A}_\mu - i g [\bold{A}_\mu,\bold{A}_\nu],
\end{equation}
contains the color electric fields
\begin{equation}
\bold{E}_j = \bold{F}_{0j} = \partial_0 \bold{A}_j - \partial_j \bold{A}_0 - i g [\bold{A}_0,\bold{A}_j]
\label{eq:definitionE}
\end{equation}
and the color magnetic fields
\begin{equation}
\bold{B}_j = - \frac{1}{2} \epsilon_{jkl} \bold{F}_{kl} = \epsilon_{jkl} \left( \partial_l \bold{A}_k - i g \bold{A}_l \bold{A}_k\right).
\label{eq:definitionB}
\end{equation}

Let us now discuss the contributions of the different fields in \eqref{eq:decompositionAspatial} and \eqref{eq:decompositionAtemporal} to the Lagrangian. We first set all fields to zero except the singlet $\sigma$. It's color electric and magnetic fields are
\begin{equation}
\begin{split}
& (E_j)_{mn} =i (\partial_0 \sigma) \;\epsilon_{jmn},\\
& (B_j)_{mn} = i (\partial_m \sigma) \; \delta_{jn} - i (\partial_n \sigma)\; \delta_{jm} - i\, g \, \sigma^2 \; \epsilon_{jmn}.
\end{split}
\end{equation}
The Lagrangian reads
\begin{equation}
\mathscr{L}_\sigma = 6(\partial_0 \sigma)^2 - 4\, \partial_m \sigma \partial_m \sigma- 6 \, g^2 \sigma^4.
\label{eq:LagrangianSigma}
\end{equation}
The field equation for $\sigma$ is given by eq.\ \eqref{eq:EOMsigma} and the homogeneous background solves eq.\ \eqref{eq:eomsigma}. Denoting by $\delta \sigma$ the small deviations from the background the quadratic action for $\delta \sigma$ becomes
\begin{equation}
\delta \mathscr{L}_{\sigma} = 6 (\partial_0 \delta \sigma)^2 - 4 \partial_m \delta \sigma \partial_m \delta \sigma - 36 \sigma^2 \delta \sigma^2 - 24 \sigma^3 \delta \sigma.
\label{eq:Lagrangiandeltasigma}
\end{equation}
Note that $\delta\sigma$ mixes with other scalar excitations as taken into account by a corresponding term in eq.\ \eqref{eq:scalarsevenLagrangian}.

We next evaluate the action for the other excitations. The contribution from color electric fields is particularly simple in Weyl gauge,
\begin{equation}
\begin{split}
\text{tr} \; \bold{E}_j \bold{E}_j = & (\partial_0 \kappa_{jmn})^2 + 6 (\partial_0 \gamma^A_{mn})^2 + 2 (\partial_0 \gamma^B_{mn})^2 \\
& + \frac{20}{3} (\partial_0 \beta^A_m)^2 + 4 (\partial_0 \beta_m^B)^2 + 6 (\partial_0 \sigma)^2.
\end{split}
\label{eq:QuadraticLagrangianColorElectricFields}
\end{equation}
Here the squares are shorthands including index summations, i.\ e.\ $(\partial_0 \gamma^A_{mn})^2 = \sum_{m,n} \partial_0 \gamma^A_{mn} \partial_0 \gamma^A_{mn}$.
In terms of the decomposition \eqref{eq:decompBeta} - \eqref{eq:decompKappa} eq.\ \eqref{eq:QuadraticLagrangianColorElectricFields} becomes
\begin{equation}
\begin{split}
\text{tr} \; \bold{E}_j \bold{E}_j = & (\partial_0 \hat \kappa_{jmn})^2 \\
&+ 3 (\partial_0 \partial_j \hat \kappa_{mn})^2 
+ 6(\partial_0 \hat \gamma^A_{mn})^2 
+ 2 (\partial_0 \hat \gamma^B_{mn})^2\\
& + \frac{8}{3} (\partial_0 \partial_m \partial_n \kappa_j)^2
+ 12 (\partial_0 \partial_j \hat \gamma^A_m)^2 
 + 4 (\partial_0 \partial_j \hat \gamma^B_m)^2 
+ \frac{20}{3} (\partial_0 \hat \beta^A_m)^2 
+ 4 (\partial_0 \hat \beta^B_m)^2 \\
& + \frac{2}{3}(\partial_0\partial_m\partial_n\partial_j \kappa)^2
+4 (\partial_0\partial_m\partial_n \gamma^A)^2
+ \frac{4}{3}(\partial_0\partial_m\partial_n \gamma^B)^2
+ \frac{20}{3}(\partial_0 \partial_m \beta^A)^2 
+ 4 (\partial_0 \partial_m \beta^B)^2 \\
&+ 6 (\partial_0 \sigma)^2.
\end{split}
\label{eq:ElectricFieldContributionFullDecomposition}
\end{equation}
The contribution from color magnetic fields is more complicated. For general $\bold{A}_j$ it reads
\begin{equation}
\begin{split}
- \text{tr} \bold{B}_j \bold{B}_j =  \text{tr} {\bigg [} & (\partial_l \bold{A}_k) (\partial_k \bold{A}_l) - (\partial_l \bold{A}_k) (\partial_l \bold{A}_k)\\
& - 2 i g \, (\partial_l \bold{A}_k) \bold{A}_k \bold{A}_l + 2 i g \, (\partial_l \bold{A}_k) \bold{A}_l \bold{A}_k \\
& - g^2 \bold{A}_l \bold{A}_k \bold{A}_k \bold{A}_l + g^2 \bold{A}_l \bold{A}_k \bold{A}_l \bold{A}_k {\bigg ]}.
\end{split}
\label{eq:colormagneticfieldstrace}
\end{equation}
In the presence of the condensate $\sigma$, Eq.\ \eqref{eq:colormagneticfieldstrace} contributes in general terms that are linear, quadratic, cubic and quartic in the remaining fields in the decomposition \eqref{eq:decompositionAspatial}. However, the requirement of rotation and discrete charge conjugation invariance leads to important restrictions. For example, a linear term must be rotation invariant and even under the discrete charge symmetry. The only possibilities are therefore the scalar fields $\delta \sigma$, $\gamma^B$ and $\beta^B$. Using Eq.\ \eqref{eq:linearTermsGeneral} gives the contribution
\begin{equation}
\mathscr{L}^{(1)} =  16 g \sigma^2 \partial_l^2 \beta^B,
\label{eq:linearbeta}
\end{equation}
in addition to the term linear in $\delta \sigma$ in \eqref{eq:Lagrangiandeltasigma}. For constant $\sigma$ eq.\ \eqref{eq:linearbeta} is a total derivative and can be dropped.

Terms following from eq.\ \eqref{eq:colormagneticfieldstrace} that are quadratic in the gluon fields different from the condensate can be obtained from a formula derived in appendix \ref{app:QuadraticTerms}.
It is clear that terms with different transformation properties under generalized rotations and charge conjugations do not mix. This holds also for the time derivative terms in \eqref{eq:ElectricFieldContributionFullDecomposition}. It is therefore useful to discuss the different representations separately. We display here only the contribution due to the divergence-less and trace-less rank-three tensor with odd charge parity,
\begin{equation}
\mathscr{L}^{(2)} = (\partial_0 \hat \kappa_{jmn})^2 -(\partial_k \hat \kappa_{jmn})^2 - 2 g^2 \sigma^2 (\hat \kappa_{jmn})^2 + 4 g\sigma\, \hat\kappa_{lmn} (\partial_j \hat \kappa_{lmk}) \epsilon_{jnk}.
\label{eq:tensor3Lagrangian}
\end{equation}
Corresponding terms for the other excitations can be found in appendix \ref{app:GaugedFixedQuadraticAction}.

\subsection{Constrained lagrangian and dispersion relations}
\begin{figure}
\begin{center}
\includegraphics[width=0.5\textwidth]{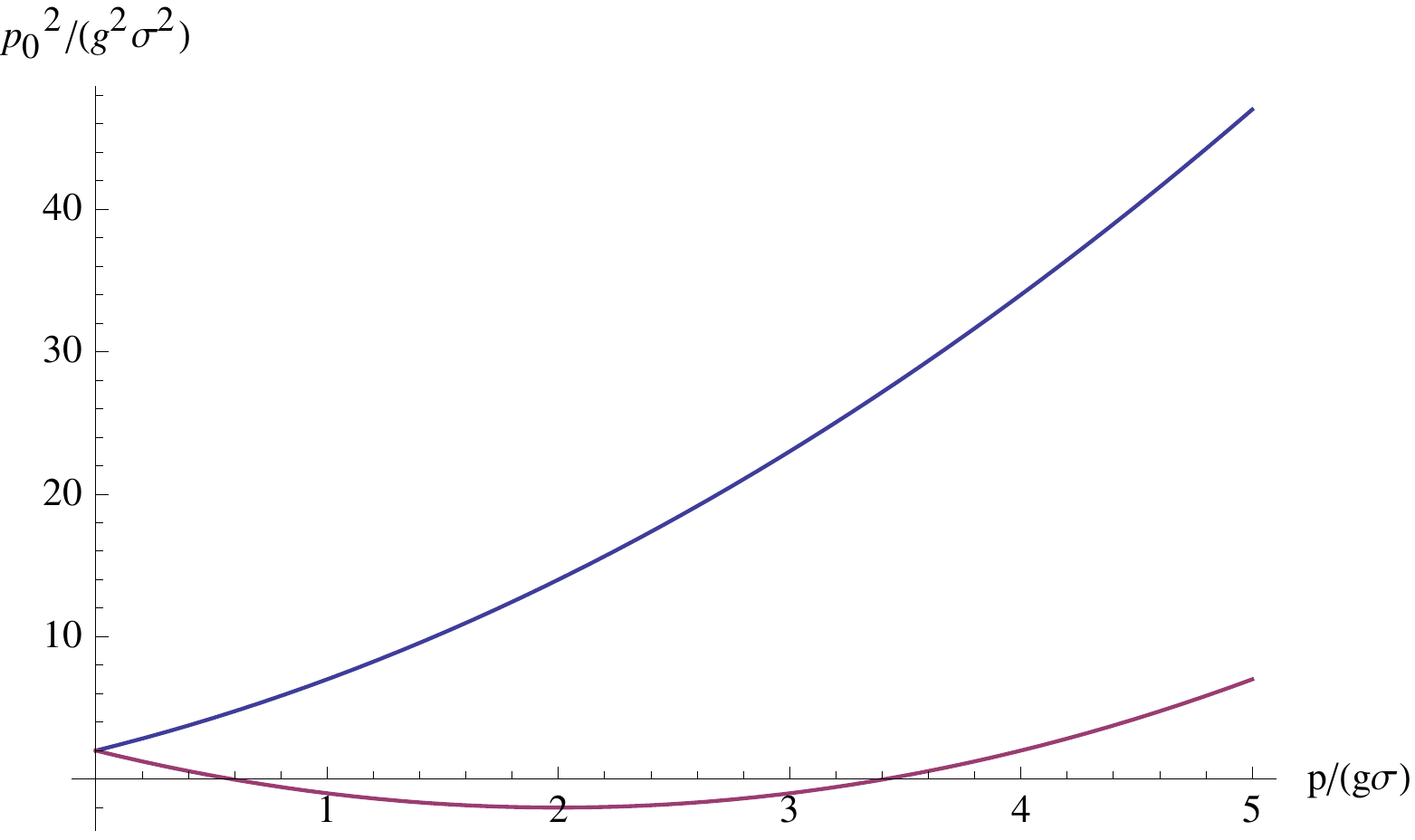}
\caption{Dispersion relations for $\hat \kappa_{jmn}$ in the presence of the condensate $\sigma$.}
\label{fig1}
\end{center}
\end{figure}

The constraints arising from the $\mu=0$ component of the field equation \eqref{eq:F1} are derived in appendix \ref{app:ConstraintEquations}. Here we display directly the contribution to the quadratic lagrangian from fields in different representations of the (modified) rotation group after the constraints are taken into account. From there the dispersion relations are computed by performing the remaining diagonalizations. The tensor of rank three $\hat \kappa_{jmn}$ is not constrained. Its contribution to the lagrangian \eqref{eq:tensor3Lagrangian} is therefore not modified. The dispersion relations for the two independent components are determined in momentum space. Choosing without loss of generality $\vec p = (0,0,p)$ one can write using the symmetry and traceless conditions
\begin{equation}
\begin{split}
& \hat\kappa_{111} = - \hat\kappa_{122} = - \hat\kappa_{221} = - \hat\kappa_{212} = a, \\
& \hat\kappa_{222} = - \hat\kappa_{211} = - \hat\kappa_{112} = - \hat\kappa_{121} = b.
\end{split}
\label{eq:kappadecompab}
\end{equation}
All other components vanish since $\hat\kappa_{jmn}$ is divergence-less, $p_j \hat \kappa_{jmn}=0$. Using \eqref{eq:kappadecompab} in \eqref{eq:tensor3Lagrangian} allows us to determine  the inverse propagator as a $2\times 2$ matrix in the space of $(a,b)$. The zeros of the determinant yield the dispersion relations which read (with $p=\sqrt{\vec p^2}$, $\sigma=\sqrt{\sigma^2}$ and $g>0$)
\begin{equation}
p_0^2 = \vec p^2 + 2 g^2 \sigma^2 + 4 p g \sigma,
\end{equation}
and
\begin{equation}
p_0^2 = \vec p^2 + 2 g^2 \sigma^2 - 4 p g \sigma,
\end{equation}
respectively. Graphical representations are shown in Fig. \ref{fig1}. 
Note that the second mode is unstable for momenta in the range
\begin{equation}
(6-4\sqrt{2}) g^2 \sigma^2 < \vec p^2 < (6+4 \sqrt{2}) g^2 \sigma^2.
\end{equation}

\begin{figure}
\begin{center}
\includegraphics[width=0.5\textwidth]{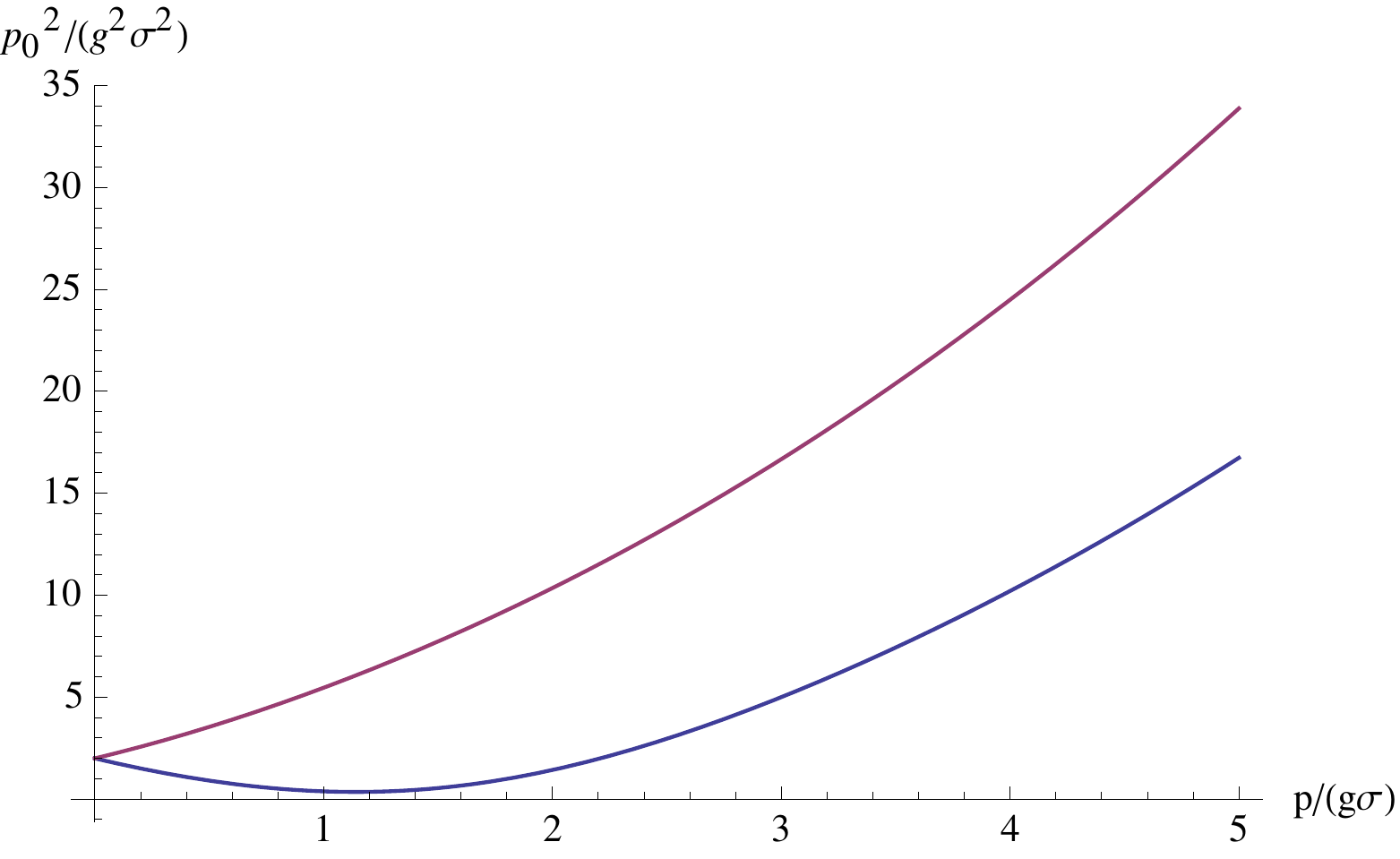}
\caption{Dispersion relations for $\hat\gamma^A_{mn}$ in the presence of the condensate $\sigma$.}
\label{fig2}
\end{center}
\end{figure}

Let us now come to the tensors of rank two with odd charge parity. The constraint in \eqref{eq:constraint1} can be used e.\ g.\ to eliminate $\hat \kappa_{mn}$ from Eq.\ \eqref{eq:tensor2oddLagrangian}. One is left with two independent modes for $\hat \gamma^A_{mn}$. Their dispersion relations are
\begin{equation}
p_0^2 = \frac{108 g^4 \sigma^4 + 180 g^3 \sigma^3 p + 108 g^2 \sigma^2 p^2 + 25 g \sigma p^3 + 5 p^4}{6(9 g^2 \sigma^2 + 3 g \sigma p + p^2)}
\end{equation}
and
\begin{equation}
p_0^2 = \frac{108 g^4 \sigma^4 - 180 g^3 \sigma^3 p + 108 g^2 \sigma^2 p^2 - 25 g \sigma p^3 + 5 p^4}{6(9 g^2 \sigma^2 - 3 g \sigma p + p^2)}.
\end{equation}
A graphical representation is shown in Fig. \ref{fig2}.
Both modes are stable.

\begin{figure}
\begin{center}
\includegraphics[width=0.5\textwidth]{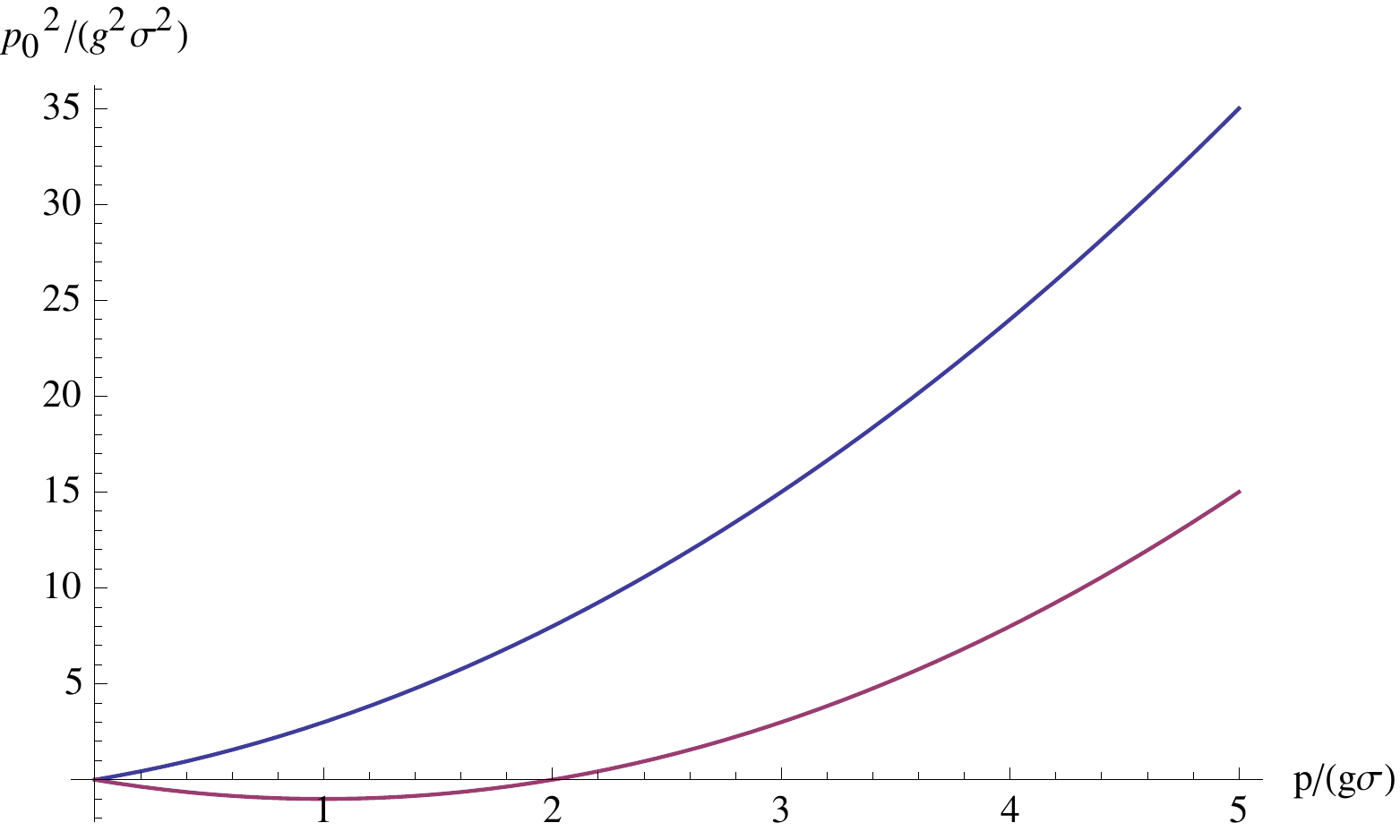}
\caption{Dispersion relations for $\hat\gamma^B_{mn}$ in the presence of the condensate $\sigma$.}
\label{fig3}
\end{center}
\end{figure}

The rank two tensors with even charge parity are not constrained. From eq.\ \eqref{eq:tensor2evenLagrangian} one obtains the dispersion relations for the two independent modes
\begin{equation}
\begin{split}
p_0^2 & = p^2 + 2 g \sigma p,\\
p_0^2 & = p^2 - 2 g \sigma p.
\end{split}
\end{equation}
The second mode is again unstable for a range of small spatial momenta. We observe the absence of a gap, i.\ e.\ $\varepsilon(\vec p=0)=0$. A graphical representation is shown in Fig. \ref{fig3}.

The vectors with odd charge parity in \eqref{eq:vectorsoddLagrangian} are constrained by \eqref{eq:constraint2} which can be used e.\ g.\ to eliminate $\hat \kappa_n$. Due to the mixing between the remaining components $\hat \gamma_m^A$ and $\hat \beta^A_m$ the dispersion relations for the four independent excitations are somewhat complicated and we do not discuss them further here. For the vectors with even charge parity the constraint in \eqref{eq:constraint4} can be used to eliminate $\hat \gamma^B_m$. Again we do not discuss the remaining independent excitations further here.

\begin{figure}
\begin{center}
\includegraphics[width=0.5\textwidth]{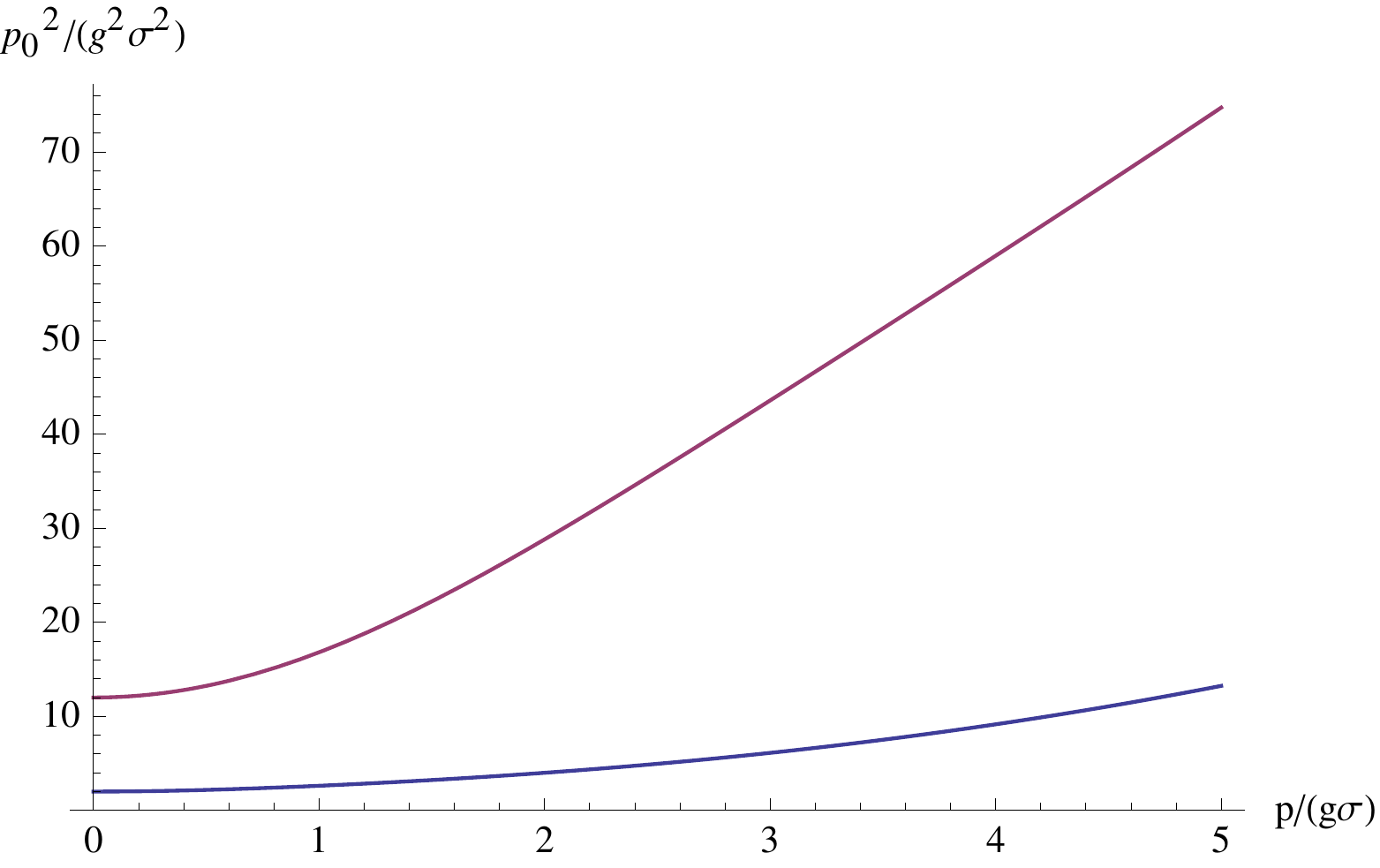}
\caption{Dispersion relations for scalar excitations with odd charge parity in the presence of the condensate $\sigma$.}
\label{fig4}
\end{center}
\end{figure}

Let us now come to scalar valued excitations with odd charge parity. With help of eq.\ \eqref{eq:constraint3} one can eliminate $\kappa$ from eq.\ \eqref{eq:scalarsoddLagrangian}. The remaining components $\gamma^A$ and $\beta^A$ still mix and the dispersion relations can be obtained from the determinant of the inverse propagator matrix in the space of $(\gamma^A,\beta^A)$,
\begin{equation}
\begin{split}
& \det
\begin{pmatrix}
4 p_0^2 p^4 + 216 g^2\sigma^2 p_0^2 p^2
- 4 p^6 - 168 g^2 \sigma^2 p^4 - 432 g^4 \sigma^4 p^2 
&, &
-72 g \sigma p_0^2 p^2 + 144 g^3 \sigma^3 p^2 \\
-72 g \sigma p_0^2 p^2  + 144 g^3 \sigma^3 p^2
&, &
\frac{92}{3} p_0^2 p^2
-20 p^4 - 128 g^2 \sigma^2 p^2
\end{pmatrix}=0.
\end{split}
\end{equation}
A graphical representation of the corresponding dispersion relations is shown in Fig. \ref{fig4}. Both modes are stable.

\begin{figure}
\begin{center}
\includegraphics[width=0.5\textwidth]{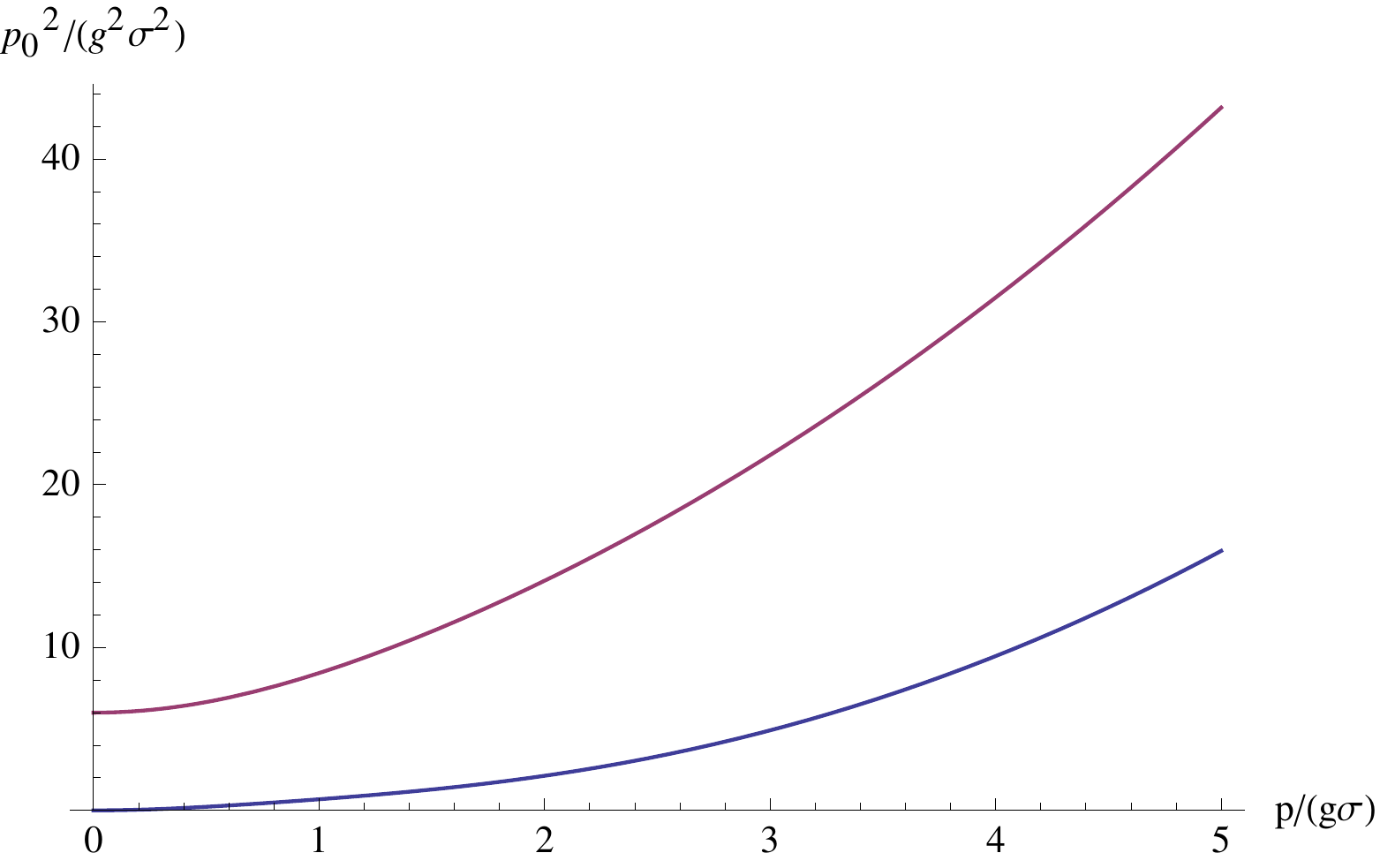}
\caption{Dispersion relations for scalar excitations with even charge parity in the presence of the condensate $\sigma$.}
\label{fig5}
\end{center}
\end{figure}

Finally, in the sector of scalar excitations with even charge parity one has to take the constraint \eqref{eq:constraint5} as well as the mixing with excitations of the scalar field $\delta\sigma$ around the constant value $\sigma$ into account. 
The dispersion relations can be obtained from the determinant of the inverse propagator matrix in the space of $(\beta^B, \delta \sigma)$,
\begin{equation}
\begin{split}
& \det
\begin{pmatrix}
4 p_0^2 p^2 + 12 g^2 \sigma^2 p_0^2 - 12 g^2 \sigma^2 p^2 - 4 p^4
&,&
- 6 g\sigma p_0^2 - 10 g\sigma p^2 \\
- 6 g\sigma p_0^2 - 10 g\sigma p^2 &,&
9 p_0^2 - 9p^2 - 36 g^2 \sigma^2
\end{pmatrix}=0.
\end{split}
\end{equation}
A graphical representation of the dispersion relations is shown in Fig.\ \ref{fig5}. We see again a gapless mode.

Comparing the different figures of this section it is apparent that the isotropic color field condensate is essentially stable with respect to homogeneous fluctuations or, more generally, fluctuations with small momenta. There is a small instability for $\hat \gamma^B_{mn}$, cf.\ fig.\ \ref{fig3}. For momenta close to zero the associated time scale for the instability is large, however. More sizeable instabilities occur for momenta of the order $(1-2) g \sigma$, cf.\ figs.\ \ref{fig2} and \ref{fig3}. As discussed above, they indicate particle production, with time scales around 1 fm/c. 
\section{Conclusions}
\label{sec:Conclusions}

The high gluon occupation numbers produced in the initial stage of a heavy ion collision may find a simple qualitative description in terms of color field condensates or gluon field condensates. The dynamics of color fields describes collective excitations and differs strongly from scattering processes involving a few coherent gluons. It provides for an efficient mechanism to achieve an approximate isotropy of the pressure on short time scales of about 0.2 fm/c. We discuss solutions of the Yang-Mills field equations that demonstrate this generic effect. They may provide a simple qualitative explanation for the findings of the numerical simulations in ref.\ \cite{Gelis:2013rba}.

This paper puts particular emphasis on the possible role of an isotropic color field condensate. Gluon fields with rotation symmetry can be realized if the rotations of the vector components are compensated by suitable gauge transformations. If such a condensate dominates, the energy momentum tensor is isotropic. The decay of the isotropic condensate into particles will preserve the isotropy of the pressure. Our stability analysis suggests that an isotropic color field condensate could play an important role in time between 0.2 fm/c and 1-2 fm/c after the collision. Its decay may be the main source of the production of gluons with isotropically distributed thermal momenta.
\begin{appendix}
\section{Decomposition of source terms}
\label{app:DecompositionSourceTerms}

We display here how an external source term for the color field transforms under modified rotations. The discussion is best done in terms of the representation
\begin{equation}
(J^\mu)_{mn} = J^{\mu a} t^a_{mn}.
\end{equation}
One can decompose $(J^0)_{mn}$ and $(J_j)_{mn}$ similarly as the gauge field in \eqref{eq:decompositionAspatial} and \eqref{eq:decompositionAtemporal},
\begin{equation}
\begin{split}
(J_j)_{mn} = & (J_\kappa)_{jmn} +(J_{\gamma^A})_{mk}\epsilon_{kjn} + (J_{\gamma^A})_{nk} \epsilon_{kjm}+ i\, (J_{\gamma^B})_{jk} \epsilon_{kmn} \\
& + ((J_{\beta^A})_m+i\,(J_{\beta^B})_m) \,\delta_{jn} + ((J_{\beta^A})_n-i\,(J_{\beta^B})_n)\, \delta_{jm} - \frac{2}{3}(J_{\beta^A})_j \delta_{mn}  + i\, J_\sigma\, \epsilon_{jmn},
\end{split}
\label{eq:decompositionJspatial}
\end{equation}
and
\begin{equation}
(J^0)_{mn}=(J_{\gamma^{C}})_{mn} + i\, (J_{\beta^{C}})_l \epsilon_{lmn}.
\label{eq:decompositionJtemporal}
\end{equation}
A generic external source breaks the modified rotation invariance. An exception is the source for the singlet field $J_\sigma$.
\section{Calculation of linear and quadratic terms}
\label{app:QuadraticTerms}

In this appendix we discuss the calculation of contributions to the Lagrangian that are quadratic in the perturbations around the color field condensate. We start from Eq.\ \eqref{eq:colormagneticfieldstrace} and insert the decomposition (we use $\epsilon_j$ for the matrix with components $\epsilon_{jmn}$)
\begin{equation}
\bold{A}_j = \tilde{\bold{A}}_j + i \sigma_0 \,\epsilon_{j}
\end{equation}
where $\tilde{\bold{A}}_j$ stands for all components in \eqref{eq:decompositionAspatial} except the constant scalar part. We drop then all terms that are not quadratic in $\tilde{\bold{A}}_j$. This gives
\begin{equation}
\begin{split}
\mathscr{L}_2 = \text{tr} {\Bigg [} & (\partial_l  \tilde{\bold{A}}_k) (\partial_k \tilde{\bold{A}}_l) - (\partial_l \tilde{\bold{A}}_k)(\partial_l \tilde{\bold{A}}_k) \\
& + g \sigma_0 {\bigg (} 2 ( \partial_l \tilde{\bold{A}}_k ) \tilde{\bold{A}}_k \bold{\epsilon}_l + 2 (\partial_l \tilde{\bold{A}}_k) \bold{\epsilon}_k \tilde{\bold{A}}_l - 2 (\partial_l \tilde{\bold{A}}_k) \tilde{\bold{A}}_l \epsilon_k - 2 (\partial_l \tilde A_k) \epsilon_l \tilde{\bold{A}}_k {\bigg )} \\
& + g^2 \sigma_0^2 {\bigg (}  
\tilde{\bold{A}}_l  \tilde{\bold{A}}_k \epsilon_k \epsilon_l - \tilde{\bold{A}}_l  \tilde{\bold{A}}_k \epsilon_l \epsilon_k
+\tilde{\bold{A}}_l   \epsilon_k \tilde{\bold{A}}_k \epsilon_l - \tilde{\bold{A}}_l \epsilon_k \tilde{\bold{A}}_l \epsilon_k \\
&\quad\quad\quad +\tilde{\bold{A}}_l   \epsilon_k \epsilon_k \tilde{\bold{A}}_l  - \tilde{\bold{A}}_l \epsilon_k \epsilon_l \tilde{\bold{A}}_k 
+\epsilon_l  \tilde{\bold{A}}_k \tilde{\bold{A}}_k  \epsilon_l   - \epsilon_l \tilde{\bold{A}}_k \tilde{\bold{A}}_l \epsilon_k  \\
&\quad\quad\quad +\epsilon_l  \tilde{\bold{A}}_k \epsilon_k \tilde{\bold{A}}_l     - \epsilon_l \tilde{\bold{A}}_k \epsilon_l \tilde{\bold{A}}_k  
+\epsilon_l  \epsilon_k \tilde{\bold{A}}_k \tilde{\bold{A}}_l     - \epsilon_l \epsilon_k \tilde{\bold{A}}_l \tilde{\bold{A}}_k
{\bigg )}{\Bigg ]}.
\end{split}
\end{equation}
Using partial integration, the cyclic property of the trace and properties of the $\epsilon$ tensor this can be reduced a bit further to
\begin{equation}
\begin{split}
\mathscr{L}_2 = & \partial_k (\tilde A_k)_{mn} \partial_l (\tilde A_l)_{nm} - \partial_l (\tilde A_k)_{mn} \partial_l (\tilde A_k)_{nm}\\
& + 2 \,g \,\sigma_0 {\Bigg [} \partial_l (\tilde A_k)_{mn} \left[ (\tilde A_k)_{ni} \epsilon_{lim} + (\tilde A_l)_{im} \epsilon_{kni} - (\tilde A_l)_{ni} \epsilon_{kim} - (\tilde A_k)_{im}  \epsilon_{lni}\right] {\Bigg ]}\\
& +g^2 \sigma_0^2 {\Bigg [}  -2 (\tilde A_l)_{mn} \left[ (\tilde A_l)_{nm} + (\tilde A_l)_{mn}\right] + 2 (\tilde A_l)_{mn} (\tilde A_m)_{nl} - 4 (\tilde A_m)_{mn} (\tilde A_k)_{nk}\\
& \quad\quad\quad\quad + (\tilde A_l)_{mn} (\tilde A_k)_{ij} \left[\epsilon_{kni} \epsilon_{ljm} + \epsilon_{lni} \epsilon_{kjm} \right] {\Bigg ]}.
\end{split}
\end{equation}
In a similar way one obtains the contribution that is linear in $\tilde {\bold A}_j$,
\begin{equation}
\begin{split}
\mathscr{L}_1 = & 2 i \partial_k (\tilde A_l)_{mn} \left[ (\partial_l \sigma) \epsilon_{knm} - (\partial_k \sigma) \epsilon_{lnm} \right]\\
& + 4 i g \sigma^2 \left[\partial_n (\tilde A_m)_{mn}  - \partial_m (\tilde A_n)_{mn} \right] \\
& - 4 i g^2 \sigma^3 (\tilde A_l)_{mn} \epsilon_{lmn}.
\end{split}
\label{eq:linearTermsGeneral}
\end{equation}
From these expressions one can obtain the specific expressions used in section \ref{sec:StabilityAnalysis}. 
\section{Gauged fixed quadratic action}
\label{app:GaugedFixedQuadraticAction}
In this appendix we display the quadratic action for the non-singlet excitations in Weyl gauge. The piece for the totally symmetric rank three tensor is already given by eq.\ \eqref{eq:tensor3Lagrangian}.
For divergence-less tensors of rank two with odd charge parity one gets
\begin{equation}
\begin{split}
\mathscr{L}^{(3)}= & 3 (\partial_0 \partial_i \hat \kappa_{mn})^2 + 6(\partial_0 \hat \gamma^A_{mn})^2 \\
& - 6 g^2 \sigma^2 (\partial_i \hat \kappa_{mn})^2 - 2 (\partial_i \partial_j \hat \kappa_{mn})^2 
+ 8 g \sigma \, (\partial_i \hat \kappa_{mn}) (\partial_i\partial_j \hat \kappa_{mk}) \epsilon_{jnk}\\
& -12 g^2 \sigma^2 (\hat \gamma^A_{mn})^2 - 2 (\partial_i \hat \gamma^A_{mn})^2 
-4 g \sigma\, \hat\gamma^A_{mn} (\partial_j \hat\gamma^A_{mk}) \epsilon_{jnk}\\
& + 4 g \sigma\, (\partial_i \hat \kappa_{mn})(\partial_i \hat \gamma^A_{mn}) - 4 (\partial_i \hat \gamma^A_{mn}) (\partial_i \partial_j \hat \kappa_{mk}) \epsilon_{jnk}.
\end{split}
\label{eq:tensor2oddLagrangian}
\end{equation}
Similarly, the contribution of rank-two tensors with even charge parity is
\begin{equation}
\mathscr{L}^{(4)}= 2(\partial_0 \hat \gamma^B_{mn})^2 - 2 (\partial_i \hat \gamma^B_{mn})^2 + 4 g \sigma \,\hat \gamma^B_{mn} (\partial_j \hat \gamma^B_{mk})\epsilon_{jnk},
\label{eq:tensor2evenLagrangian}
\end{equation}
while the contribution from vectors with odd charge parity reads
\begin{equation}
\begin{split}
\mathscr{L}^{(5)}= & \frac{8}{3}(\partial_0\partial_i\partial_j \hat \kappa_m)^2 
+12(\partial_0\partial_i \hat \gamma^A_m)^2 
+ \frac{20}{3} (\partial_0 \hat \beta^A_m)^2\\
& -\frac{16}{9} (\partial_i \partial_j \partial_m \kappa_n)^2 
-\frac{16}{3} (\partial_i\partial_j \kappa_m)^2 g^2 \sigma^2 
+\frac{32}{9} g \sigma (\partial_i \partial_m \hat \kappa_n) (\partial_i \partial_m \partial_j \hat \kappa_k) \epsilon_{jnk}\\
& -10 (\partial_i \partial_j \hat \gamma^A_m)^2 
-24 g^2 \sigma^2 (\partial_i \hat \gamma^A_m)^2 
- 4 g \sigma (\partial_i \hat \gamma^A_n) (\partial_i \partial_j \hat \gamma^A_k) \epsilon_{jnk}\\
& -\frac{14}{3} (\partial_i \hat \beta^A_m)^2  
-80 g^2 \sigma^2 (\hat \beta^A_m)^2 
+ 20g \sigma \hat \beta^A_n (\partial_j \hat \beta^A_k) \epsilon_{jnk}\\
& + \frac{16}{3} g \sigma (\partial_i \partial_j \hat\kappa_m) (\partial_i \partial_j \hat\gamma^A_m)
- \frac{8}{3} (\partial_i \hat \kappa_n) (\partial_i  \partial_j \hat \gamma^A_k) \epsilon_{jnk}\\
& -\frac{8}{3} (\partial_i\partial_j\hat\kappa_m)(\partial_i \partial_j \hat\beta_m)\\
& -48 g \sigma (\partial_i \hat \gamma^A_m) (\partial_i \hat \beta^A_m) 
+ 4 (\partial_i \hat \gamma^A_n) (\partial_i \partial_j \hat \beta^A_k) \epsilon_{jnk},
\end{split}
\label{eq:vectorsoddLagrangian}
\end{equation}
and the contribution from vectors with even charge parity is given by
\begin{equation}
\begin{split}
\mathscr{L}^{(6)}= & 4 (\partial_0\partial_j \hat \gamma^B_m)^2 + 4 (\partial_0 \hat \beta^B_m)^2 \\
& - 2 (\partial_i \partial_j \hat \gamma^B_m)^2 + 4 g \sigma (\partial_i \gamma^B_n) (\partial_i \partial_j \gamma^B_k) \epsilon_{jnk}\\
& -2 (\partial_i \hat \beta^B_m)^2 - 8 g^2 \sigma^2 (\hat \beta^B_m)^2 -4 g \sigma \hat\beta^B_n (\partial_j \beta^B_k) \epsilon_{jnk}\\
& - 4 (\partial_i \hat \gamma^B_n)(\partial_i \partial_j \hat \beta^B_k) \epsilon_{jnk}.
\end{split}
\end{equation}
Finally, the scalar fields with odd charge parity contribute
\begin{equation}
\begin{split}
\mathscr{L}^{(7)}= & \frac{2}{3} (\partial_0 \partial_i\partial_j \partial_m \kappa)^2 
+4 (\partial_0\partial_i \partial_j \gamma^A)^2 
+ \frac{20}{3}(\partial_0 \partial_i \beta^A)^2 \\
& -\frac{4}{9} (\partial_i\partial_j\partial_m \partial_n \kappa)^2
-\frac{4}{3} g^2\sigma^2 (\partial_i\partial_j \partial_m \kappa)^2\\
& - 4 (\partial_i \partial_j \partial_m \gamma^A)^2
- 8 g^2\sigma^2 (\partial_i \partial_j \gamma^A)^2\\
& - 4 (\partial_i \partial_j \beta^A)^2
- 80 g^2 \sigma^2 (\partial_i \beta^A)^2 \\
& + \frac{8}{9} g \sigma (\partial_i \partial_j \partial_m \kappa) (\partial_i \partial_j \partial_m \gamma^A)\\
& - 32 g \sigma (\partial_i\partial_j \gamma^A) (\partial_i\partial_j \beta^A)
,
\end{split}
\label{eq:scalarsoddLagrangian}
\end{equation}
and the scalar fields with even charge parity yield
\begin{equation}
\begin{split}
\mathscr{L}^{(8)}= & \frac{4}{3} (\partial_0 \partial_i \partial_j \gamma^B)^2  
+ 4(\partial_0\partial_i \beta^B)^2\\
& -\frac{4}{9} (\partial_i \partial_j \partial_m \gamma^B)^2 \\
& - 4 (\partial_i\partial_j \beta^B)^2
-8 g^2 \sigma^2 (\partial_i \beta^B)^2\\
& - \frac{8}{3} (\partial_j^2 \delta \sigma) (\partial_l^2 \gamma^B) 
+ 32 g \sigma (\delta \sigma) (\partial_l^2 \beta^B)
.
\end{split}
\label{eq:scalarsevenLagrangian}
\end{equation}
Terms quadratic in $\delta \sigma$ have already been taken into account in eq.\ \eqref{eq:Lagrangiandeltasigma}.
\section{Constraint equations}
\label{app:ConstraintEquations}

We have neglected the temporal component of the gauge field $(A_0)_{mn}$ so far by choosing Weyl gauge. There is, however, one important set of equations we did not take into account so far. It is obtained from variation of the Lagrangian with respect to $(A_0)_{mn}$,
\begin{equation}
\partial_j (E_j)_{mn} - i g (A_j)_{mk} (E_j)_{kn} + i g (E_j)_{mk} (A_j)_{kn} = D_j (E_j)_{mn} = 0.
\label{eq:constraintGeneral}
\end{equation}
We use now again Weyl gauge where one has the relation $(E_j)_{mn} = \partial_0 (A_j)_{mn}$. Moreover, we write the gluon field as
\begin{equation}
(A_j)_{mn} = i \sigma \epsilon_{jmn} + (\tilde A_j)_{mn},
\end{equation}
where the condensate $\sigma$ is assumed to be constant in time and space. Expanding eq.\ \eqref{eq:constraintGeneral} to linear order in $(\tilde A_j)_{mn}$ gives
\begin{equation}
\partial_0 \left[ \partial_j (\tilde A_j)_{mn} + g \sigma \epsilon_{jmk} (\tilde A_j)_{kn} - g \sigma \epsilon_{jkn} (\tilde A_j)_{mk} \right] = 0.
\end{equation}
This can be divided into a tensor type constraint
\begin{equation}
\begin{split}
& \partial_0 {\bigg [}\partial_j\, \kappa_{jmn}+\epsilon_{kjn} (\partial_j \gamma^A_{mk}) + \epsilon_{kjm} (\partial_j \gamma^A_{nk})+ \partial_m \beta^A_n + \partial_n \beta^A_m  - \frac{2}{3} \partial_j \beta^A_j \,\delta_{mn} - 6 \, g \sigma  \, \gamma^A_{mn} {\bigg ]} = 0,
\end{split}
\label{eq:constraint1}
\end{equation}
and a vector type constraint
\begin{equation}
\partial_0 {\bigg [} \partial_k  \gamma^B_{jk}  
+ \epsilon_{jmn} \partial_n  \beta^B_m + \partial_j \delta \sigma  - 2 \, g\sigma \, \beta^B_j {\bigg ]}= 0.
\label{eq:constraint2}
\end{equation}

In terms of the decomposition \eqref{eq:decompBeta}, \eqref{eq:decompGamma} and \eqref{eq:decompKappa} the constraint \eqref{eq:constraint1} corresponds to the tensor type condition
\begin{equation}
\partial_0 \left[ \partial_j^2 \hat \kappa_{mn} + \partial_j \hat \gamma^A_{mk} \epsilon_{jnk} + \partial_j \hat \gamma^A_{nk} \epsilon_{jmk} - 6 g \sigma \hat \gamma^A_{mn} \right] = 0,
\label{eq:constraint1a}
\end{equation}
the vector type condition
\begin{equation}
\partial_0 \partial_m \left[ \frac{2}{3} \partial_j^2 \hat \kappa_n + \partial_j \hat \gamma^A_k \epsilon_{jnk} + \hat \beta^A_n - 6 g \sigma\, \hat \gamma^A_n \right] = 0,
\label{eq:constraint2a}
\end{equation}
and the scalar type condition
\begin{equation}
\partial_0 \partial_m \partial_n \left[ \frac{1}{3} \partial_j^2 \kappa + 2 \beta^A - 6 g \sigma \gamma^A \right] = 0.
\label{eq:constraint3}
\end{equation}
Similarly, the constraint \eqref{eq:constraint2} corresponds to the vector type condition
\begin{equation}
\partial_0 \left[ \partial_k^2 \hat \gamma^B_m - \partial_j \hat \beta^B_k \epsilon_{jmk} - 2 g \sigma\, \hat \beta^B_m \right] = 0
\label{eq:constraint4}
\end{equation}
and the scalar type condition
\begin{equation}
\partial_0 \partial_m \left[\frac{2}{3}\partial_k^2 \gamma^B + \delta\sigma - 2 g \sigma \beta^B \right] = 0.
\label{eq:constraint5}
\end{equation}
One can use the remaining gauge freedom within Weyl gauge to make sure that the quantities within the brackets in eqns.\ \eqref{eq:constraint1a} - \eqref{eq:constraint5} are not only conserved but vanish identically. One sees that this allows to express the fields $\hat \kappa_{mn}$, $\hat \kappa_m$, $\kappa$, $\hat\gamma^B_m$ and $\gamma^B$ in terms of other, independent, fields. 

\end{appendix}

\end{document}